\documentclass[longauth]{aa}  

\usepackage{csvsimple}
\usepackage{xcolor}
\usepackage{float}
\usepackage{stfloats}
\usepackage{graphicx}
\usepackage{graphbox}
\usepackage{chngcntr}
\usepackage{amsmath}
\usepackage{amssymb}
\usepackage{mathtools}
\usepackage{multirow}
\usepackage{verbatim}
\usepackage[toc]{appendix}
\usepackage{soul}
\usepackage[varg]{txfonts}
\usepackage{JournalAbb}
\usepackage{natbib}
\usepackage[colorlinks,citecolor=blue,urlcolor=blue,bookmarks=false,hypertexnames=true]{hyperref} 
\bibpunct{(}{)}{;}{a}{}{,}
\makeatletter
\renewcommand*\aa@pageof{, page \thepage{} of \pageref*{LastPage}}
\makeatother

\defcitealias{Bourdin2017}{B17}
\defcitealias{Kozmanyan2019}{K19}

\begin{document} 

    \title{CHEX-MATE: Relationship between X-ray and millimetre inferences of galaxy cluster temperature profiles}
  
    \titlerunning{CHEX-MATE: Relation of X-ray and mm inferences of cluster temperatures}

    \author{
        F. De Luca\thanks{\email{federico.deluca@roma2.infn.it}}\inst{\ref{UniR2},\ref{INFN-R2}}
        \and
        H. Bourdin\inst{\ref{UniR2},\ref{INFN-R2}}
        \and
        P. Mazzotta\inst{\ref{UniR2},\ref{INFN-R2}}
        \and
        G. Luzzi\inst{\ref{ASI}}
        \and
        M. G. Campitiello\inst{\ref{Argonne}}
        \and
        M. De Petris\inst{\ref{UniR1}}
        \and
        D. Eckert\inst{\ref{UniGE}}
        \and
        S. Ettori\inst{\ref{INAF-oas},\ref{INFN-bo}}
        \and
        A. Ferragamo\inst{\ref{UniNa}}
        \and
        W. Forman \inst{\ref{CfA}}
        \and
        M. Gaspari\inst{\ref{UniMore}}
        \and
        F. Gastaldello\inst{\ref{INAF-IASF}}
        \and
        S. Ghizzardi \inst{\ref{INAF-IASF}}
        \and
        M. Gitti\inst{\ref{UniBo-DIFA}, \ref{IRA}}
        \and
        S.~T. Kay\inst{\ref{Jodrell}}
        \and
        J. Kim\inst{\ref{KAIST}}
        \and
        L. Lovisari\inst{\ref{CfA},\ref{INAF-IASF}}
        \and
        J.F. Macías-Pérez \inst{\ref{LPSC}}
        \and
        B. J. Maughan\inst{\ref{UBhhw}}
        \and
        M. Muñoz-Echeverría\inst{\ref{IRAP}}
        \and
        F. Oppizzi\inst{\ref{INFN-Ge}}
        \and
        E. Pointecouteau\inst{\ref{IRAP}}
        \and
        G.~W. Pratt\inst{\ref{Saclay}}
        \and
        E. Rasia\inst{\ref{INAF-oat}, \ref{IFPU}, \ref{Umich}}
        \and
        M. Rossetti\inst{\ref{INAF-IASF}}
        \and
        H. Saxena\inst{\ref{Caltech}}
        \and
        J. Sayers\inst{\ref{Caltech}}
        \and
        M. Sereno\inst{\ref{INAF-oas},\ref{INFN-bo}}
    }

    \institute{
        Dipartimento di Fisica, Università di Roma ‘Tor Vergata’, Via della Ricerca Scientifica 1, I-00133 Roma, Italy.\label{UniR2}
        \and
        INFN, Sezione di Roma ‘Tor Vergata’, Via della Ricerca Scientifica, 1, 00133, Roma, Italy.\label{INFN-R2}
        \and
        Space Science Data Center, Italian Space Agency, via del Politecnico, 00133, Roma, Italy\label{ASI}
        \and
        High Energy Physics Division, Argonne National Laboratory, Lemont, IL 60439, USA\label{Argonne}
        \and
        Dipartimento di Fisica, Sapienza Università di Roma, Piazzale Aldo Moro 5, I-00185 Rome, Italy\label{UniR1}
        \and
        Department of Astronomy, University of Geneva, Ch. d’Ecogia 16, CH-1290 Versoix, Switzerland\label{UniGE}
        \and
        INAF - Osservatorio di Astrofisica e Scienza dello Spazio di Bologna, via Piero Gobetti 93/3, I-40129 Bologna, Italy\label{INAF-oas}
        \and
        INFN, Sezione di Bologna, viale Berti Pichat 6/2, I-40127 Bologna, Italy\label{INFN-bo}
        \and
        Dipartimento di Fisica ’E. Pancini’, Università degli Studi di Napoli Federico II, Via Cinthia, 21, I-80126 Napoli, Italy\label{UniNa}
        \and
        Center for Astrophysics | Harvard \& Smithsonian, 60 Garden Street, Cambridge, MA 02138, USA\label{CfA}
        \and
        Department of Physics, Informatics and Mathematics, University of Modena and Reggio Emilia, 41125 Modena, Italy\label{UniMore}
        \and
        INAF - IASF Milano, via A. Corti 12, I-20133, Milano, Italy\label{INAF-IASF}
        \and
        Dipartimento di Fisica e Astronomia (DIFA), Alma Mater Studiorum - Università di Bologna, via Gobetti 93/2, 40129 Bologna, Italy\label{UniBo-DIFA}
        \and
        Istituto Nazionale di Astrofisica – Istituto di Radioastronomia (IRA), via Gobetti 101, 40129 Bologna, Italy\label{IRA}
        \and
        Jodrell Bank Centre for Astrophysics, Department of Physics and Astronomy, The University of Manchester, Manchester M13 9PL, UK\label{Jodrell}
        \and
        Department of Physics, Korea Advanced Institute of Science and Technology (KAIST), 291 Daehak-ro, Yuseong-gu, Daejeon 34141, Republic of Korea\label{KAIST}
        \and
        Université Grenoble Alpes, CNRS, LPSC-IN2P3, 53, avenue des Martyrs, 38000 Grenoble, France\label{LPSC}
        \and
        HH Wills Physics Laboratory, University of Bristol, Bristol, BS8 1TL, UK\label{UBhhw}
        \and
        IRAP, CNRS, Université de Toulouse, CNES, UT3-UPS, Toulouse, France\label{IRAP}
        \and
        INFN, Sezione di Genova, Via Dodecaneso 33, 16146, Genova, Italy\label{INFN-Ge}
        \and
        Université Paris-Saclay, Université Paris Cité, CEA, CNRS, AIM, 91191 Gif-sur-Yvette, France\label{Saclay}
        \and
        INAF - Osservatorio Astronomico di Trieste, via Tiepolo 11, I-34131, Trieste, Italy\label{INAF-oat}
        \and
        IFPU, Institute for Fundamental Physics of the Universe, Via Beirut 2, 34014 Trieste, Italy\label{IFPU}
        \and
        Department of Physics; University of Michigan, Ann Arbor, MI 48109, USA\label{Umich}
        \and
        California Institute of Technology, 1200 East California Boulevard, Pasadena, California, USA\label{Caltech}
    }

    
    \date{Received Month Day, Year; accepted Month Day, Year}


    \abstract{Thermodynamic profiles from X-ray and millimetre observations of galaxy clusters are often compared under the simplifying assumptions of smooth, spherically symmetric intracluster medium. These approximations lead to expected discrepancies in the inferred profiles, which can provide insights about the cluster structure or cosmology. Motivated by this, we present a joint XMM-\textit{Newton} and \textit{Planck} analysis of 116 CHEX-MATE clusters to measure $\eta_T = T_X/T_{SZ,X}$, the ratio between spectroscopic X-ray temperatures and a temperature proxy derived from Sunyaev-Zel'dovich (SZ) pressures and X-ray densities. We considered relativistic corrections to the thermal SZ signal and implemented X-ray absorption by Galactic molecular hydrogen. The $\eta_T$ distribution has a mean of $1.01 \pm 0.03$, with average changes of $8.1\%$ and $2.7\%$ when relativistic corrections and molecular hydrogen absorption are not included, respectively. The $\eta_T$ distribution is positively skewed, with the scatter mostly affected by cluster morphology: relaxed clusters are closer to unity and less scattered than mixed and disturbed systems. We find little or no correlation with redshift, mass, or temperature.}


    \maketitle
    
\section{Introduction} \label{sec:Intro}
    
    Clusters of galaxies are complex, multicomponent systems observable over a large part of the electromagnetic spectrum. In optical and near-IR wavelengths, clusters appear as overdense regions in the galaxy distribution over the sky. However, most of the mass is not hosted in galaxies. Dark matter dominates the total mass budget, and the vast majority of baryonic content \citep[at least 80\%, e.g.][]{Gonzalez2013, Chiu2018, Eckert2019} resides in the intracluster medium (ICM). The ICM is a highly ionised plasma, heated to high temperatures (about $10^7$--$10^8 \,{\rm K}$) during the hierarchical cluster growth \citep{Kravtsov2012}. It is observable at X-ray frequencies,  with a combination of bremsstrahlung and metal line emissions \citep[e.g.][]{Mitchell1976, Serlemitsos1977, Sarazin1986, Bohringer2010}. In physical units\footnote{If $\Sigma_X$ is expressed in terms of photon fluxes (i.e. without accounting for photon energies, $\rm counts\, cm^{-2} arcmin^{-2} sr^{-1}$), the redshift dimming factor in Eq.~\eqref{eq:X_sb} becomes $(1+z)^{-3}$ \citep{Reese2000}.} (${\rm erg\, s^{-1}cm^{-2}sr^{-1}}$), the X-ray surface brightness, within an arbitrary energy band, is given by:
    \begin{equation}
        \Sigma_X(r)=\frac{1}{4\pi(1+z)^4}\int [n_p n_e](r)\Lambda_X(T,Z) \, dl,
        \label{eq:X_sb}
    \end{equation}    
    where $z$ is the cluster redshift, $n_p$, $n_e$ are the proton and electron number densities, and $\Lambda_X$ is the X-ray cooling function for bremsstrahlung and metal line emission. 
    \\
    At millimetre wavelengths, the free electrons in the ICM are also responsible for the inverse Compton scattering distortion of the Cosmic Microwave Background (CMB), the Sunyaev-Zel'dovich effect \citep{Sunyaev1972,Sunyaev1980}. Considering the non-relativistic Kompaneets approximation \citep{Kompaneets1957}, we can analytically describe the spectral shape of the thermal SZ (tSZ) effect with $Y_0(x)$, while the amplitude of the signal ($I_{SZ}$) is proportional to the Compton parameter $y$ \citep{Sunyaev1972, Birkinshaw1999, Carlstrom2002, Mroczkowski2019}:
    \begin{align}
        &I_{SZ}(x) = \frac{\Delta T_{CMB}(x)}{T_{CMB}} = y \cdot Y_0(x) = y \left[x\coth(x/2)-4\right],
        \label{eq:I_y}  \\
        &y = \int \dfrac{kT_e}{m_e c^2} d\tau = \dfrac{\sigma_T}{m_e c^2}\int n_e kT_e dl = \dfrac{\sigma_T}{m_e c^2}\int P_e \, dl,
        \label{eq:y}
    \end{align}
    where $x = h\nu/kT$ is the dimensionless frequency, $h$ and $k$ the Planck and Boltzmann constants, $\sigma_T$ and $\tau$ the Thomson cross section and line-of-sight optical depth, and $m_e$, $T_e$, $P_e$ the electron mass, temperature, and pressure. 
    \\
    Relativistic electrons and magnetic fields in the ICM can also produce diffuse radio emission (synchrotron), tracing the past cluster growth via accretion shocks, merging events, or turbulence (\citeauthor{Brunetti2009} \citeyear{Brunetti2009}; \citeauthor{Balboni2025} \citeyear{Balboni2025}, or for reviews, see \citeauthor{Brunetti2014} \citeyear{Brunetti2014}; \citeauthor{Weeren2019} \citeyear{Weeren2019}).
    
    Therefore, observations of galaxy clusters provide peculiar insight into a wide range of astrophysical processes, from galactic to cosmological scales. In fact, cluster formation and evolution are strongly affected by the underlying cosmological framework \citep[see e.g.][]{Voit2005, Allen2011}. Constraints on cosmological models can be derived from the spatial distribution and number density of cluster samples from dedicated X-ray, SZ, or optical surveys \citep[as done, for example, in][]{Vikhlinin2009, Planck2015XXIV, Pacaud2018, Costanzi2021, Ghirardini2024}, or from their sizes, using them as cosmological rulers \citep[e.g.][]{Cowie1978, Silk1978, Cavaliere1979, Carlstrom2002, Reese2002}. However, cosmological results may be biased by modelling assumptions such as spherical symmetry, hydrostatic equilibrium, and ICM inhomogeneities due to substructures, mergers, shocks, and turbulence \citep[e.g.][]{Gaspari2014, Pratt2019, Pearce2020, Ansarifard2020, Gianfagna2021}. A multiwavelength approach is then essential to comprehensively describe cluster properties and extract reliable cosmological information. 
    
    In this context, the SZ signal probes the line-of-sight (LOS) integrated electron pressure (Eq.~\ref{eq:y}) and is unaffected by redshift dimming. On the other hand, the X-ray surface brightness depends on the integrated square of the electron density for a given chemical composition (Eq.~\ref{eq:X_sb}), making it more sensitive to the denser central regions. In addition, X-ray spectroscopy can be used to directly measure ICM temperatures and metallicity. X-ray and SZ observations are thus highly complementary and can be combined to extract thermodynamical properties, especially when one observable is not accessible \citep[such as temperature for low S/N observations or for high-redshift clusters, see e.g.][]{Pointecouteau2002, Kitayama2004, Adam2017, Mastromarino2024}. Beyond thermodynamics, with joint X-ray and SZ observations (or even combining them with lensing data), it is possible to constrain the triaxial shape of clusters \citep{Sereno2018, Kim2024, Chappuis2025, Gavidia2025}. An effective tracer of such effects is given by $\eta$, defined as the normalisation ratio of the pressure profiles derived from X-ray and SZ observations:
    \begin{equation}
        \eta = P_X / P_{SZ} = b_n \cdot \mathcal{C} \cdot \mathcal{B} \,,
        \label{eq:eta_def}
    \end{equation}
    or, equivalently, by $\eta_T$, when defined from temperature profiles. Deviations from unity can be used as hints of systematic effects \citep[][hereafter \citetalias{Kozmanyan2019}]{Kozmanyan2019}. With this formalism, $\mathcal{B}$ accounts for deviations from the assumed cluster model, such as gas inhomogeneities, substructures, or asphericity, while $\mathcal{C}$ encodes cosmological biases \citepalias[e.g. uncertainties in the angular diameter distance or in the primordial chemical composition, see][for further details]{Kozmanyan2019}. Any additional bias, independent of the previous, is denoted by $b_n$ in Eq.~\eqref{eq:eta_def}, and may be related to residual systematics from instrumental calibration or profile modelling and fitting (e.g. assuming a non-relativistic SZ model). 
    
    Several studies have measured $\eta$ to constrain cosmological parameters. With 61 clusters observed with XMM-\textit{Newton} and \textit{Planck}, \citet[][hereafter \citetalias{Bourdin2017}]{Bourdin2017} derived a median $\eta_T$ of $1.01^{+0.02}_{-0.04}$, which \citetalias{Kozmanyan2019} used to estimate $H_0 = 67 \pm 3$ km\,s$^{-1}$\,Mpc$^{-1}$. Analysing the X-COP sample, \citet{Ghirardini2019} found $\eta = 0.9624 \pm 0.0013$ (r.m.s. $0.08$), while \citet{Ettori2020} showed that the inferred helium abundance strongly depends on the assumed Hubble constant. Including relativistic corrections to the SZ spectra, \citet{Wan2021} analysed 14 dynamically relaxed and massive clusters observed with Chandra, \textit{Planck}, and Bolocam, whose data yield a median $\eta$ of $1.14$ (denoted as $\mathcal{R}$ in their work) and inferring $H_0 = 67.3^{+21.3}_{-13.3}$ km\,s$^{-1}$\,Mpc$^{-1}$. 
    
    Following this series of investigations, we present a joint SZ and X-ray analysis of cluster observations to study systematic mismatches in the reconstruction of temperature profiles ($\eta_T$), to be used for cosmological studies. We extend the methodology of \citetalias{Bourdin2017,Kozmanyan2019} and apply it to a new large sample of galaxy clusters, presented for the first time by the \citet{CHEX-MATE}. In particular, we include more sources of systematics in the thermodynamical analysis, such as a more robust quantification of the X-ray soft proton contamination, the additional absorption from molecular hydrogen in the X-ray analysis, and the relativistic corrections in the tSZ effect. This paper is structured as follows. The sample of galaxy clusters is presented in Section~\ref{sec:Sample}. In Sections~\ref{sec:Method} and \ref{sec:Results}, we describe the methodology and the results, respectively. Finally, our findings are summarised in Section~\ref{sec:Conclusions}. Hereafter, we assume a flat cosmological model with cold dark matter and a cosmological constant associated with dark energy (the $\Lambda$CDM model), with $H_0= \rm 70\, km\, s^{-1} Mpc^{-1}$, $\Omega_m=0.3$, and $\Omega_\Lambda=0.7$. The characteristic radii, masses, or pressures of clusters are expressed in terms of overdensities ($\Delta$) relative to the critical density of the Universe, $\rho_c (z)$, evaluated at the cluster redshift. Thus, for a density contrast of $\Delta=500$: $M_{500} = (4\pi/3)\, 500\, \rho_c(z)\, R_{500}^3$.
    
\section{The galaxy cluster sample} \label{sec:Sample}

    \begin{figure}
    	\includegraphics[width=\columnwidth]{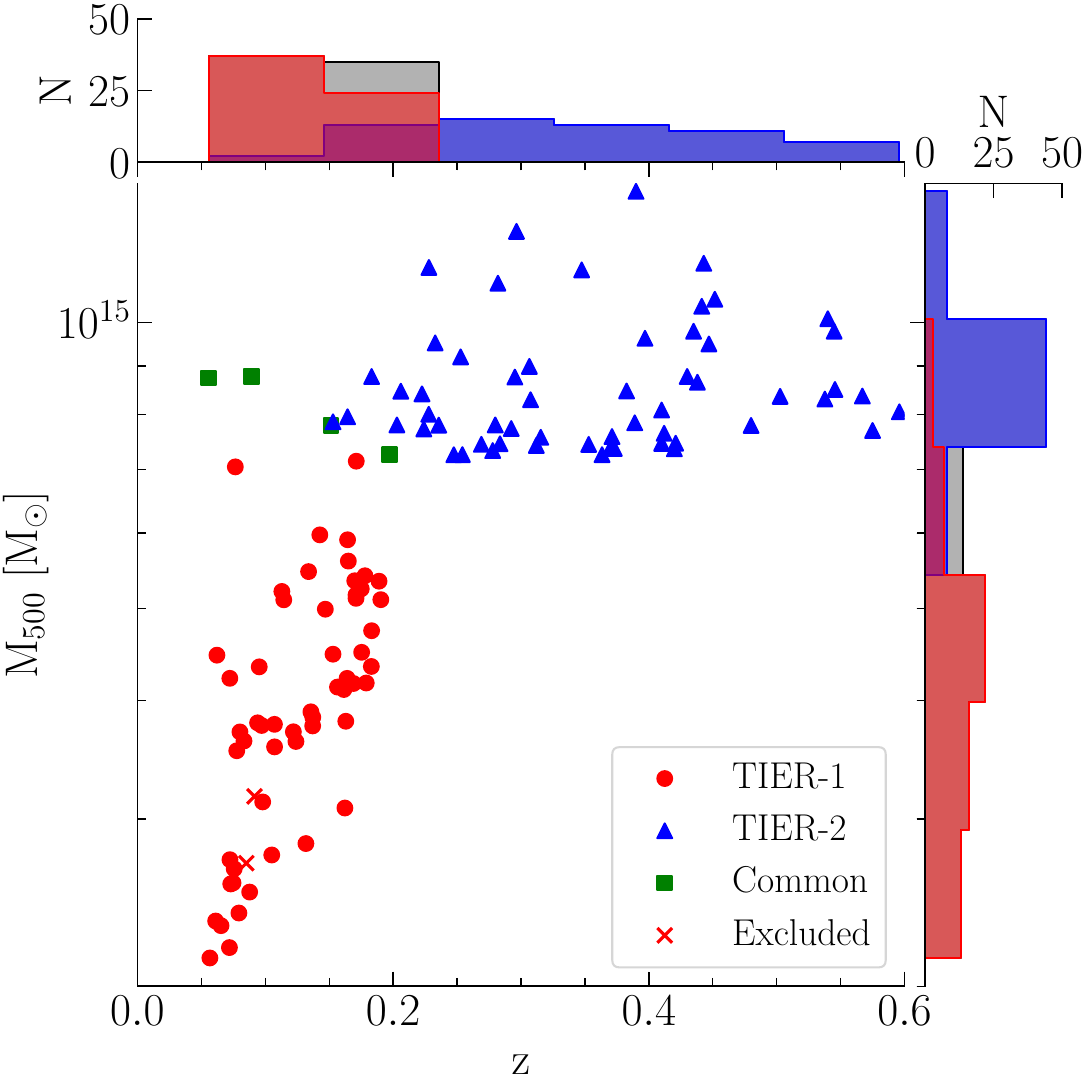}
        \caption{Mass and redshift distribution of the CHEX-MATE sample (grey). Tier-1 and Tier-2 clusters are shown as red circles and blue triangles, respectively, while common clusters as green squares. Excluded clusters from the analysis are shown with red crosses.}
        \label{fig:Sample}
    \end{figure}

    The Cluster HEritage project\footnote{\url{http://xmm-heritage.oas.inaf.it/}} with XMM-\textit{Newton} -- Mass Assembly and Thermodynamics at the Endpoint of structure formation (CHEX-MATE) is a three-megasecond, multi-year XMM-\textit{Newton} programme of X-ray observations of galaxy clusters. It is a minimally biased, signal-to-noise-limited ($\rm S/N>6.5$) sample of 118 galaxy clusters built from the second \textit{Planck} all-sky SZ catalogue \citep[PSZ2, ][]{Planck2015XXVII}. The CHEX-MATE sample is composed of two subsamples: Tier-1, collecting low-redshift clusters, and Tier-2, including some of the most massive systems in the Universe. In Fig.~\ref{fig:Sample}, we show the mass and redshift distributions of the sample, where the Tier-1 and Tier-2 clusters are marked with red circles and blue triangles, respectively. The four objects in common between the two subsamples are shown as green squares. For cluster masses and redshifts, we use the values presented in \citet{CHEX-MATE}, and derived from the \textit{Planck} SZ signal \citep{Planck2015XXVII} using the MMF3 algorithm \citep{Melin2006, Planck2013XXIX} and the scaling relation $Y_{SZ}-M_{500}$, without corrections for hydrostatic bias. The complete sample covers the redshift range $0.05<z<0.6$ (median: $0.18$) and the masses in the interval $M_{500}\in (2,14)\times10^{14}{\rm M_\odot}$ (median: $M_{500}=7.2\times10^{14}{\rm M_\odot}$). We refer the reader to the introductory paper \citep{CHEX-MATE} for a more detailed description of the CHEX-MATE project, the galaxy cluster sample definition, the observing strategy, and the possible outcomes. Unfortunately, this sample includes two clusters with problematic contributions from foreground or background sources. For PSZ2G283.91+73.87, the X-ray observation is highly contaminated by the foreground emission of the Virgo cluster outskirts, as shown in App.~\ref{app:gallery}. PSZ2G028.63+50.15 instead presents an X-ray double-halo structure, with the addition of a third background cluster located close to the cluster centre \citep[see also][]{Schellenberger2022}. These background and foreground contributions, if not properly modelled, could significantly affect the estimation of the thermodynamic radial profiles. Thus, we excluded these two systems from our analysis.
    
\section{Method} \label{sec:Method}

   In this section, we describe the method to characterise the X-ray and SZ signals of galaxy clusters coming from the XMM-\textit{Newton} and \textit{Planck} space telescopes, with a discussion about the modelling of backgrounds and foregrounds. We note that the strategy followed in this work consists of a revision of the method developed in \citetalias{Bourdin2017} and differs from the pipeline used for other CHEX-MATE products, such as for the X-ray surface brightness or temperature profiles presented in \citet{Bartalucci2023, Rossetti2024}. We tested the consistency of the X-ray spectroscopy analyses in App.~\ref{app:T_comparison}, with 28 clusters of the DR1 subsample (built to be representative of the overall CHEX-MATE sample) presented in \citet{Rossetti2024}. In summary, the two pipelines return concordant results concerning the spectroscopic temperatures from X-ray observations, in the radial range explored in this work.
   
\subsection{X-ray analysis} \label{ssec:X-ray ana}

    The X-ray observations analysed in this work were performed with the three EPIC cameras \citep[MOS1, MOS2, and PN,][]{Turner2001, Struder2001} on board the XMM-\textit{Newton} satellite. The Observation Data Files (ODF), publicly available on the XMM-\textit{Newton} Science Archive\footnote{\url{http://nxsa.esac.esa.int/nxsa-web/\#home}}, are firstly preprocessed using the Science Analysis System\footnote{\url{https://www.cosmos.esa.int/web/xmm-newton/sas}} (SAS) pipeline, version 18.0.0. To generate calibrated event lists, the ODFs are preprocessed using SAS \textsc{emchain} and \textsc{epchain} tools. For multiple observations of the same targets, the most cleaned data available in the repository are collected and analysed for our study. We clean data from high-background periods and solar flares, filtering the light curve profile with a temporal wavelet analysis, as in \citet{Bourdin2008} and \citet{Bourdin2013}. In particular, we remove from the datasets all the detected periods in which we have more than $3\sigma$ deviations from the light curve distributions in the high ($10-12\, \rm keV$) or soft ($1-5\, \rm keV$) X-ray bands. MOS1 or MOS2 CCDs with anomalous count rates \citep[e.g. see][]{Kuntz2008} are also excluded.

\subsubsection{X-ray background and foreground models} \label{ssec:X-ray background}

    The X-ray background is characterised by instrumental and astrophysical components. The instrumental background derives from the electronic noise and the interactions of high energy particles with the telescope. These interactions produce a quiescent particle background (QPB) characterised by a flat spectrum with fluorescence lines \citep{Katayama2004, Kuntz2008, Snowden2008, Marelli2021}. We estimate the QPB component after the light curve filtering of the calibrated event lists (see Sect.~\ref{ssec:X-ray ana}). The normalisation of the QPB spectrum is estimated using the energy band where it is dominant: $10$--$12\, \rm keV$ for MOS and $12$--$14\, \rm keV$ for PN. To reduce any possible contamination from the clusters, we restrict our analysis to data located beyond $1.5R_{500}$ from the X-ray centre. 
    \\
    Considering the observational strategy of the CHEX-MATE sample, for most of the pointings, the field of view (FOV) subtends a region around the clusters larger than $3R_{500}$. Offset observations are also present in the catalogue for background modelling \citep{CHEX-MATE}. For the few cases where such regions are not available, we consider a circular annulus of radii $13\arcmin$--$15\arcmin$, centred on the aim point. In this work, we consider as cluster centre the X-ray peak, defined as the coordinates of the maximum of a wavelet-denoised X-ray surface brightness map, in the energy band $[0.5, 2.5]\, \rm keV$. We use a cubic B-spline wavelet \citep[for a review, see][]{Starck2002, Starck2009} to denoise the X-ray images corrected for point sources, background noise, and spatial variations of the effective areas. Specifically, wavelet coefficients of a MultiScale Variance Stabilized Transform \citep{Starck2009} are soft-thresholded\footnote{For a given threshold $\lambda$ and a value of wavelet coefficient $d$, soft-thresholding is defined as: $D(d|\lambda) = sgn(d)(0, |d|- \lambda)_+$ \citep{Donoho1995}.} at a $4\sigma$ confidence level to suppress Poisson noise.
    
    The astrophysical component originates from the foreground emission of our Galaxy, the presence of point sources, and unresolved X-ray sources that constitute the Cosmic X-ray Background. The foreground Galactic emission can be described using two spectral components: one associated with the thermal emission of the Local Hot Bubble and the second with thermal ‘transabsorption’ emission from the galactic halo \citep{Kuntz2000, Lumb2002}. We model the spectral and spatial features of these components as in \citet{Bourdin2013}, considering the same region used for the particle background. Point sources in the field of view are detected and masked using \textsc{SExtractor} \citep{SExtractor}. The results of the point-source masking are visually inspected for possible other unidentified sources (e.g. other extended sources not related with the CHEX-MATE clusters) or incorrect detections.
    
    In addition to QPB and astrophysical backgrounds, a residual focused contribution originally attributed to soft protons \citep[SP, but whose origin is still unclear, e.g.][]{Salvetti2017, Gastaldello2022, Fioretti2024, Mineo2024} can affect X-ray observations even after light curve filtering \citep[see, for example,][]{DeLuca2004, Kuntz2008, Snowden2008, Leccardi2008, Lovisari2019}. To consider any residual SP contribution, we model the SP spectra as a power law with index $-0.6$ \citep[as in][]{Rossetti2024}. SP modelling is one of the differences between our method and the CHEX-MATE analysis, which considers a more detailed physical and predictive modelling \citep[for more details, see][]{Rossetti2024}. Despite its simplicity, our model returns consistent results with the standard CHEX-MATE pipeline in the regions of interest (especially within $R_{500}$), as discussed in App.~\ref{app:T_comparison}.
    
\subsubsection{X-ray cluster spectral model and radial profiles} \label{ssec:X-ray cl sp model}

    Cluster emission is modelled with the APEC\footnote{\url{http://www.atomdb.org/index.php}} spectral library for thermal bremsstrahlung and line emissions from metals in the ICM \citep{Smith2001}. The resulting spectrum is then corrected for the absorption of X-ray photons due to the Galactic Inter-Stellar Medium. In particular, we use the absorption cross-sections from \citet{Verner1996}, the element abundance table of \citet{Asplund2009}, and the total (atomic plus molecular) hydrogen column density estimated in \citet{Bourdin2023}, with the exception of PSZ2G263.68-22.55 where the column density is directly estimated from the X-ray spectrum, as in \citet[][Appendix~A]{Oppizzi2022}.
    
    To calculate $\eta_T$ from the X-ray and SZ data, we deproject the ICM profiles using a forward approach, adopting the analytic profiles of \citet{Vikhlinin2006} for density and temperature:
    \begin{align}
        [n_p n_e](r) = &\, \frac{n_0^2(r/r_c)^{-\alpha}}{\left[1+(r/r_s)^\gamma\right]^{\epsilon/\gamma} \left[1+(r/r_{c})^2\right]^{3\beta_1-\alpha/2}} + \nonumber
        \\
        & + \frac{n_{0,2}^2}{\left[1+(r/r_{c,2})^2\right]^{3\beta_2}}, \label{eq:3D_Xprofne}
        \\
        T(r)= &\, T_0 \frac{x+T_{min}/T_0}{x+1} \frac{(r/r_t)^{-a}}{\left[1+(r/r_t)^b\right]^{c/b}}.
        \label{eq:3D_Xprofte}
    \end{align}
    For the density profile, $n_0$ and $n_{0,2}$ are the normalisations of two $\beta$-models \citep{Cavaliere1976, Cavaliere1978}, with slopes $\beta_1$ and $\beta_2$, characteristic radii $r_c$ and $r_{c,2}$, and index $\alpha$ for the central power law cusp. The temperature profile is modelled as a broken power law with a characteristic scale $r_t$, featuring slopes $a$, $b$, and $c$ to describe the inner, transition, and outer regions. The model also includes a term for possible temperature declines for central cooling regions, expressed in terms of the typical cooling scale $x=(r/r_{cool})^{a_{cool}}$ (and where $a_{cool}$ sets the slope of the central profile) and with normalisation $T_{min}$. The overall temperature normalisation factor is parametrised with $T_0$.
    
    The deprojected density and temperature profiles are jointly extracted by fitting the surface brightness (Eq.~\ref{eq:X_sb}) and the spectroscopic temperature to account for the small dependences of the surface brightness on the temperature. In particular, to compare the observed temperature with the proposed model in the fit, we consider the spectroscopic-like temperature introduced in \citet{Mazzotta2004}:
    \begin{equation}
        T_{sl}=\frac{\int w T dl}{\int w \,dl},
        \label{eq:T_splike}
    \end{equation}
    where $w = n_e^2/T^{3/4}$. We note that all spectroscopic temperatures towards the cluster cores (i.e. where the metallicity is higher) are within the validity regime of the proposed weighting scheme \citep[i.e. $2$--$3\, \rm keV$, see][]{Mazzotta2004}. By passing from the 3D radial models to projected templates, we consider the effect of the XMM-\textit{Newton} point-spread function \citep[using the EPIC PSF model developed by][]{Ghizzardi2001, Ghizzardi2002} in the surface brightness template and the absorption model described above, with a constant cluster metal abundance of $0.3\,Z_\odot$ and the {\it Planck} primordial helium abundance \citep[Eq.~80b,][]{Planck2018VI}. As a result, we have a particle mean weight equal to $\mu=0.592$ and $n_p/n_e=0.859$ as the average proton-to-electron density.
    
    The fit of the profiles is achieved considering a least-squares Levenberg–Marquardt minimizer \citep{Markwardt2009}, while uncertainties are estimated by performing $500$ parametric bootstrap realisations of the observed profiles over the measurement errors. For the surface brightness fit, we consider an energy band of $[0.5, 2.5]\, \rm keV$, and $50$ logarithmically equally spaced annuli between the X-ray peak and $R_{500}$. For the temperature, we fit the cluster spectra in twelve annuli within $R_{500}$: four bins inside $0.15R_{500}$, seven logarithmically spaced between $0.15$ and $0.8 R_{500}$, and one from $0.8$ to $R_{500}$. We consider the $0.3$--$12.1\, \rm keV$ energy band to estimate the cluster temperature and metallicity.
    
    An exception to this procedure is made for PSZ2G339.63-69.34 (the Phoenix cluster), where the AGN signal makes the estimation of the cluster spectrum non-trivial for XMM-\textit{Newton} EPIC cameras. The central AGN emission, moderately obscured but dominant above $\sim 2\, {\rm keV}$ \citep{Tozzi2015, McDonald2019}, cannot be easily separated from the ICM due to the small apparent cluster size ($R_{500}=2.85 \arcmin$), its bright cool core, and the XMM-\textit{Newton} point spread function (half-energy width at the aimpoint is $\sim 15\,\arcsec$). As a result, the strong AGN signal contaminates the XMM-\textit{Newton} observations up to the cluster outskirts. To mitigate this contamination, we restrict the surface brightness and temperature analysis to the energy range where the cluster emission dominates (below $2\,{\rm keV}$), as in \citet{Oppizzi2022}.
    
\subsection{Millimetre analysis} \label{ssec:SZ analysis}

    In this section, we summarise the main steps for the millimetre data analysis. For a more accurate description of the method, we refer to \citetalias{Bourdin2017}, or \citet{Oppizzi2022} where this method is applied to retrieve the non-relativistic SZ signal from a joint analysis of \textit{Planck} and the South Pole Telescope (SPT) data on six CHEX-MATE galaxy clusters. In this work, we use the \textit{Planck}-HFI data from the second public data release\footnote{\url{http://pla.esac.esa.int/pla/\#home}} (PR2), considering the full 30-month mission dataset \citep{planck2015I}. In particular, 
    we select $17\degr \times 17\degr$ cluster-centred patches from the six HFI all-sky maps \citep[Nside 2048, nominal frequencies: $100$, $143$, $217$, $353$, $545$, $857\,\rm GHz$ and Gaussian FWHM: $9.69\arcmin$, $7.30\arcmin$, $5.02\arcmin$, $4.94\arcmin$, $4.83\arcmin$, $4.64\arcmin$, respectively; see][]{Planck2015VII} using the {\sc gnomview} tool of HEALPix\footnote{\url{http://healpix.sf.net}}. This results in $1024 \times 1024$ images with $1\arcmin$ pixel resolution. With respect to \citetalias{Bourdin2017}, we update the removal of point sources in HFI maps masking all the PCCS2 and PCCS2E point sources \citep{Planck2015XXVI} located more than $2.5R_{500}$ from the cluster centres. We also filter out instrumental offsets and anisotropies with scales much larger than cluster sizes, mainly of CMB and Galactic Thermal Dust (GTD) origin. Specifically, we apply a third-order B-spline high-pass filtering of width of $1\degr$ and $15\arcmin$ for nearby ($z<0.48$) and distant clusters ($z \ge 0.48$), respectively. Secondly, we estimate the spatial and spectral template of the background and foreground signals. 
    
\subsubsection{Millimetre background and foreground models} \label{ssec:mm background}
    
    Our model for backgrounds and foregrounds includes the contributions of the CMB, GTD, and an additional corrective term to account for the cluster thermal dust emission (CTD), as observed by stacked frequency map analysis \citep[e.g. see][]{Planck2015XXIII, Planck2016XLIII}. For the GTD component, we model the spatial template using the HFI $857 \,{\rm GHz}$ channel. In particular, we clean the map from noise by performing a wavelet coefficient thresholding of an isotropic undecimated wavelet transform of the image \citep{Starck2007}. For this, we use a $3\sigma$ threshold matching the power spectrum and variance of 1000 noise simulations realised from differences in the half-ring datasets. The dust Spectral Energy Distribution (SED) is modelled as suggested by \citet{Meisner2015}, with a two-temperature ($T_1$ and $T_2$) modified black body emission:
    \begin{equation}
        s_{GTD}(\nu)=\left[ \frac{f_1q_1}{q_2}\left(\frac{\nu}{\nu_0}\right)^{\beta_{d,1}}B_\nu(T_1)+(1-f_1)\left(\frac{\nu}{\nu_0}\right)^{\beta_{d,2}}B_\nu(T_2)\right],
        \label{eq:SED_dust}
    \end{equation}
    where $B_\nu$ is the Planck function, $\beta_{d,1}$ and $\beta_{d,2}$ are the corresponding power law indices, $f_1$ the cold component fraction, and $q_1$, $q_2$ the ratios of far infrared emission to optical absorption. 
    \\
    Since the GTD spatial model is modelled using the $857 \rm\, GHz$ channel, it must include any thermal dust contributions of Galactic and cluster origin. Thus, we consider a corrective term for the CTD SED that becomes null at $857 \rm\, GHz$ \citepalias[see Eq.~14 of][]{Bourdin2017}. For the spatial template of CTD, we consider the results presented in \citet{Planck2015XXIII}. In particular, we use a Navarro-Frenk-White \citep[NFW;][]{NFW1997} density profile ($\rho_{NFW}$) for the cluster dust distribution with $c_{500}=1$. The SED of the CTD is instead modelled as a single-temperature modified black body, with a fixed spectral index ($\beta_{d, CTD}=1.5$) and with a temperature corrected for the cluster redshift: $T_{CTD} = (1+z)^{-1} 24.2 {\rm K}$, as estimated in \citet{Planck2016XLIII}:
    \begin{align}
        s_{CTD}(\nu) & = \nu\,^{\beta_{d, CTD}} B_\nu(T_{CTD}), \\
        \Delta s_{CTD}(\nu) & = \left[ \frac{s_{CTD}(\nu)}{s_{CTD}(857 \rm GHz)} - \frac{s_{GTD}(\nu)}{s_{GTD}(857 \rm GHz)} \right], \\
        I_{CTD} (\nu) & = \tau_{CTD} \Delta s_{CTD}(\nu) \int \rho_{NFW} (r) \, dl,
        \label{eq:I_CTD}
    \end{align}
    and where $\tau_{CTD}$ is a dimensionless normalisation factor for $I_{CTD}$.
    \\
    Similarly to the GTD, the spatial template of the black body CMB signal is modelled considering the $217 \rm\, GHz$ channel but subtracted by the GTD template and properly denoised using the same $3\sigma$ wavelet thresholding method used for the GTD.
    
    CMB and GTD models are jointly fitted considering the HFI data within a circular annulus ($7$--$12\,R_{500}$) around the X-ray cluster peak, where the cluster emission is negligible. The fit is performed through $\chi^2$ minimisation, with $f_1$, the CMB and GTD normalisations as free parameters. For GTD temperatures, we use, for $T_2$, the temperature map of the hot dust component from the joint \textit{Planck}-IRAS analysis of \citet{Meisner2015} and consider a power-law relation for $T_1$ \citep[assuming a thermal equilibrium of the dust grains with the interstellar radiation field; for more detail see Eq.~14 of][]{Finkbeiner1999}. The other GTD parameters, which describe the properties of grain absorption and emission (and should not vary significantly considering different lines of sight), are fixed to the average all sky values of \citet{Meisner2015}. For CTD, since it originates within the cluster, we fit this component together with the SZ signal as detailed in Sect.~\ref{ssec:joint-fit}.

\subsubsection{SZ modelling and cluster radial profiles} \label{ssec:SZ model}

    For the tSZ, Eq.~\eqref{eq:I_y} is valid only in the classic limit, and there are no exact analytical solutions in the presence of relativistic electrons. However, at typical ICM temperatures ($\sim 5 \,\rm keV$), relativistic electrons may have a non-negligible impact on the tSZ amplitude and spectral shape \citep[][]{Rephaeli1995b, Hurier2016, Erler2018, Remazeilles2018, Perrott2024, Kuhn2025}. Relativistic corrections introduce explicit temperature dependence in the SZ spectrum \citep[which can be exploited to measure the ICM temperature, e.g.][]{Coulton2024arX, Remazeilles2024, Vaughan2025} and redistribute tSZ power towards higher frequencies, reducing the peak-to-trough amplitude at fixed $y$ (see App.~\ref{app:rSZ}, Fig.~\ref{fig:rSZ_example}). This affects the inferred Compton-$y$ and pressure profiles when using a non-relativistic model. Thus, relativistic corrections to the Kompaneets expression have been deduced using different approaches, generally providing approximate analytical expansions in terms of dimensionless gas temperature $\theta_e = kT_e/(m_ec^2)$ \citep[e.g.][]{Wright1979, Rephaeli1995a, Rephaeli1995b, Challinor1998, Itoh1998, Nozawa1998, Pointecouteau1998, Chluba2012}.
    
    In this work, we consider both classical (cSZ; Eq.~\ref{eq:I_y}) and relativistic (rSZ) approaches to quantify systematic mismatches in the SZ measurement. In particular, we used as a baseline the analytical approximation for the tSZ of \citet{Nozawa1998, Nozawa2006}, already applied by \citet{Luzzi2022} to constrain the CMB temperature scaling with redshift using galaxy clusters. However, this model is valid under the assumption of an isothermal ICM temperature distribution. To include any effect due to the gas temperature distribution, we expand the analytical approximation of \citet{Nozawa2006} considering a temperature central moment expansion, similarly to \citet{Prokhorov2012, Chluba2013}, as detailed in App.~\ref{app:rSZ}. As noted by \citet{Lee2020} and \citet{Kay2024}, higher order corrections become relevant at larger scales. Thus, in our relativistic tSZ model (along the LOS and at a given frequency), we consider temperature moments up to the fourth order:
    \begin{equation}
        I_{SZ}(x, \theta_e) = y(n_e, \theta_e) \cdot g(x,\bar{\theta}_e, \sigma_e^2, \gamma_e, \kappa_e), 
        \label{eq:I_rSZ}
    \end{equation}
    where $\bar{\theta}_e$, $\sigma_e^2$, $\gamma_e$, $\kappa_e$ are respectively the mean, variance, skewness, and kurtosis of the temperature distribution. For their estimate in the SZ spectrum $g(x)$, we use at each radius the fitted temperature (and density) distribution on the X-ray data (Eq.~\ref{eq:3D_Xprofte}). 

    Considering Eqs.~\eqref{eq:y} and~\eqref{eq:I_rSZ}, the Compton parameter $y$ for tSZ is modelled with the analytical gNFW profile proposed by \citet{Nagai2007} for the cluster pressure:
    \begin{equation}
        \frac{P_e(r)}{P_{500}} = \frac{P_0}{(r/r_s)^\gamma \left[1+(r/r_s)^\alpha\right]^{(\beta-\gamma)/\alpha}},
        \label{eq:3D_yprof}
    \end{equation}
    where $r_s=R_{500}/c_{500}$ is the typical scale of the profile expressed in terms of $R_{500}$, and $c_{500}$ is the concentration parameter. The parameters $\alpha$, $\beta$, and $\gamma$ are the slopes for the intermediate, outer, and inner regions, respectively. $P_0$ is the normalisation of the pressure profile. Finally, the profiles are expressed in terms of their characteristic pressure $P_{500}$ as predicted by a self-similar model \citep{Nagai2007} and defined as in \citet{Arnaud2010}:
    \begin{equation}
        P_{500}=1.65\times10^{-3}E(z)^{8/3} \left[ \frac{ M_{500}}{3\times10^{14} M_{\odot}}\right]^{2/3} \, {\rm keV\, cm^{-3}},
        \label{eq:P500}
    \end{equation}
    where, for a flat universe and assuming a negligible radiation term, $E(z)^2=\Omega_M(1+z)^3+\Omega_\Lambda$. The SZ pressure template is fitted using the SZ cluster signal, $I^d_{SZ}(\nu)$, obtained by subtracting the foreground and background components described in Sect.~\ref{ssec:mm background} from the HFI denoised data:
    \begin{equation}
        I^d_{SZ}(\nu) = I_{HFI}(\nu) - I_{CMB}(\nu) - I_{GTD}(\nu) - I_{CTD}(\nu).
        \label{eq:I_SZ}
    \end{equation}
    The modelling of each SED also includes corrections for the spectral and spatial (considering Gaussian beams) response and colour or unit corrections between the HFI channels (for a further description, see \citealp{planck2015I}, \citetalias{Bourdin2017}). Moreover, the same high-pass filtering used for the data (see Sect.~\ref{ssec:mm background}) is applied to the proposed model. The SZ signal is evaluated within $5R_{500}$ of the X-ray peak, using 15 radial bins for all clusters, except for the four systems with $R_{500} \lesssim 3\arcmin$, where 10 bins were adopted.
    
    In addition to the tSZ, another contribution to the overall SZ signal comes from the cluster bulk motion, i.e. from the kinematic SZ (kSZ) effect. Although the implemented pipeline admits possible contributions from kSZ \citep[and with relativistic correction of][]{Nozawa2006}, this component is not included in our model for the \textit{Planck} data. In fact, measurements of the peculiar motion of clusters from redshift surveys are complicated, since they require independent and accurate estimates of the cluster distances. Moreover, at the predicted cluster peculiar velocities, the kSZ is expected to contribute less \citep[e.g. about one order of magnitude less than tSZ,][]{Mroczkowski2019} to the overall SZ signal, and its spectral behaviour follows the CMB one. Thus, proper detections of the kSZ for single clusters require a large coverage of the SZ spectra (mainly to disentangle the tSZ from the kSZ) and a sufficiently high angular resolution and sensitivity to separate the SZ signals from the contaminates \citep[for some recent studies about the kSZ, see][and reference therein]{Planck2016XXXVII, Adam2017kSZ, Sayers2019}.
    
\subsection{XMM-\textit{Newton} and \textit{Planck} joint fit} \label{ssec:joint-fit}
    
    From the thermodynamical templates of the ICM density (Eq.~\ref{eq:3D_Xprofne}), temperature (Eq.~\ref{eq:3D_Xprofte}), and pressure (Eq.~\ref{eq:3D_yprof}), we derive the systematic mismatch parameter $\eta$ (Eq.~\ref{eq:eta_def}) from spectroscopic temperature comparisons, following the definition of \citetalias{Bourdin2017}. With the SZ-derived pressure and the X-ray density, we can compute a 3D temperature template (under the assumption of the ideal gas law) as $kT_e^{SZ,X}(r) = P^{SZ}_e(r)/n^X_e(r)$. This template is then converted to a projected spectroscopic estimate according to Eq.~\eqref{eq:T_splike}:
    \begin{equation}
        T_{SZ,X} = \eta_T \cdot T_{sl}^{SZ,X} = \eta_T \frac{\int w T_e^{SZ,X} dl}{\int w \, dl} ,
        \label{eq:eta_T}
    \end{equation}
    and compared with the measured X-ray spectroscopic temperatures (i.e. within $R_{500}$, see Sect.~\ref{ssec:X-ray cl sp model}). As discussed in \citet{Wan2021}, this joint fitting approach weights more the data in radial ranges simultaneously probed by both X-ray and millimetre instruments. With current X-ray telescopes, the signal reaches the background level at $\sim0.8R_{500}$, and the \textit{Planck} resolution ($\sim5 {\rm arcmin}$) limits the reconstruction of the inner regions, inside $\sim0.33R_{500}$ (considering our binsize for most of the clusters). Thus, $\eta_T$ is mostly constrained by data coming from intermediate regions (about $0.33-0.8R_{500}$). 
    
    With this formalism, any potential bias in the temperature estimates is encapsulated in the $\eta_T$ parameter, which, similarly to Eq.~\eqref{eq:eta_def}, may arise from cosmological assumptions, cluster modelling, or cross-calibration mismatches \citepalias[see, for example, Appendix A of][for an analytical study of the dependence of $\eta_T$ on these biases]{Kozmanyan2019}. Confidence intervals and best values are calculated with an MCMC joint fit of the X-ray and SZ data, with log-likelihood:
    \begin{align}
        & \ln{\mathcal{L}}=\ln{\mathcal{L}_{SZ}}+\ln{\mathcal{L}_X} 
        \label{eq:like_tot}
        \\
        & \ln{\mathcal{L}_{SZ}(\vec{\theta})}\propto \left[ \vec{I}_{SZ}^d - \vec{I}_{SZ} \left(\vec{\theta} \right) \right]^T C^{-1}_{SZ} \left[ \vec{I}_{SZ}^d - \vec{I}_{SZ} \left(\vec{\theta} \right) \right],
        \label{eq:like_SZ}
        \\
        & \ln{\mathcal{L}_{X}(\vec{\theta})}\propto \left[ \vec{T}_{X} - \vec{T}_{SZ,X} \left(\vec{\theta} \right) \right]^T C^{-1}_{X} \left[ \vec{T}_{X, j} - \vec{T}_{SZ,X} \left(\vec{\theta} \right) \right],
        \label{eq:like_X}
    \end{align}
    where $\vec{\theta}=[\eta_T,\, P_0,\, \alpha,\, \beta,\, \gamma,\, c_{500},\, \tau_{CTD}]$ is the vector of the model parameters. In Eq.~\eqref{eq:like_SZ}, $\vec{I}_{SZ}^d$ is a vector where we stack the observed radial SZ signals estimated in the HFI channels, $\vec{I}_{SZ}$ is the corresponding model, and $C_{SZ}$ is the covariance matrix for the SZ data. As in \citet{Oppizzi2022}, we compute the covariance matrix for the SZ data as the sample covariance of 1000 profiles (for each HFI channel), estimated outside the region where we fit the cluster signal. We introduce this modification to the original \citetalias{Bourdin2017} method to account for possible correlated residuals between the HFI channels due to our modelling, which has common CMB and GTD spatial templates. In Eq.~\eqref{eq:like_X}, $\vec{T}_{SZ,X}$ is the vector with projected spectroscopic-like temperatures (Eq.~\ref{eq:eta_T}), to be compared with the measured X-ray ones $\vec{T}_{X}$, and $C_{X}$ is a diagonal covariance matrix with the uncertainties on $\vec{T}_{X}$. For the MCMC setup, we use a Metropolis-Hastings sampler with five chains, initialised from random values within the prior ranges. The free parameters (and their respective priors) are $\eta_T\sim\mathcal{U}(0.25,6)$, $\tau_{CTD}\sim\mathcal{U}(0,\tau_{max})$ (where $\tau_{max} = 10$ if a preliminary ML fit returns zero for this component and $\tau_{max} = 5\tau_{ML}$ otherwise), and the SZ pressure parameters $P_0\sim\mathcal{U}(0,50)$, $\alpha\sim\mathcal{U}(0,7)$, and $\beta\sim\mathcal{U}(0,25)$.
    
    Due to the limited spatial resolution of \textit{Planck} (our bin size for most clusters is $\sim0.33R_{500}$), we fixed the other parameters ($\gamma = 0.31$, $c_{500} = 1.18$) to the universal pressure profile values of \citet{Arnaud2010} to avoid strong (and non-linear) degeneracies between the slopes and $c_{500}$ (especially when fitting individual clusters), leading to poor convergence of the chains \citep[as also noted by][]{Pointecouteau2021} or non-physical values. With this modelling, we can constrain the pressure profiles at the “larger scale” with \textit{Planck} while, for the behaviour (i.e. the normalisation) of the profiles within $R_{500}$, we use X-ray data (in combination with the proposed SZ pressure in the MCMC run). We note that in pressure fitting, the $P_{500}$ normalisations are fixed considering the MMF3 masses. As noted in \citet{Munoz2025} studying a subsample of CHEX-MATE clusters, such a modelling introduces a correlation between the profile shape and the adopted mass. The impact of this correlation for the complete CHEX-MATE sample will be investigated in future work of the collaboration.

\section{Results} \label{sec:Results}

\subsection{Effect of relativistic corrections in the SZ signal and pressure estimates} \label{ssec:rSZ-cl P diff}

    \begin{figure*}
    	\includegraphics[width=2\columnwidth]{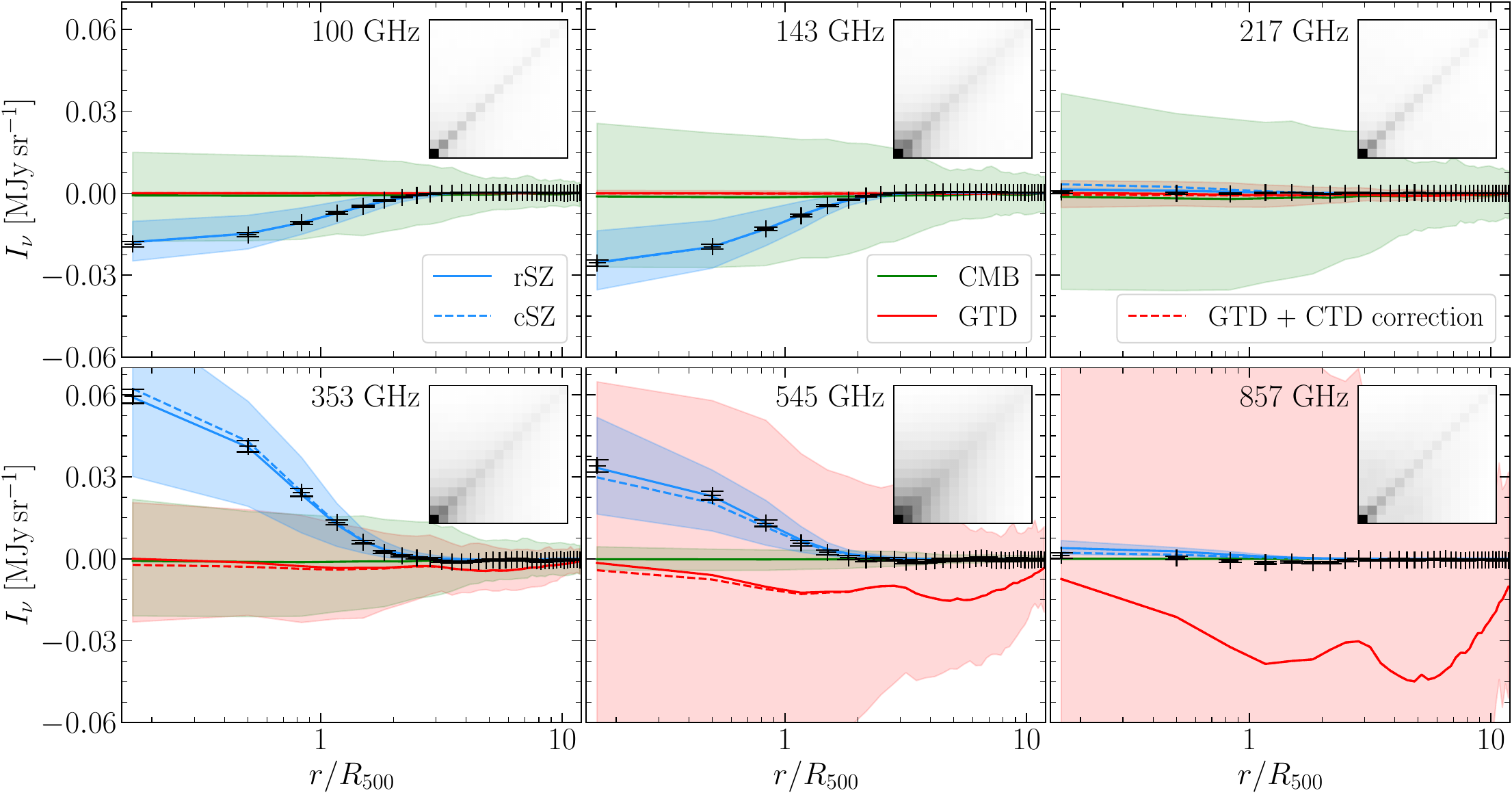}
        \caption{Stacked mean radial profiles in the \textit{Planck} HFI channels towards 112 CHEX-MATE clusters. Black crosses with error bars show the observed mean SZ signal and its standard uncertainty. The CMB and GTD mean models are shown as solid green and red lines, respectively. The red dash-dotted line marks the dust component after the CTD correction. Solid and dashed blue lines represent the mean rSZ and cSZ models. Shaded areas cover the $16$th–$84$th percentiles intervals. The normalised mean covariance matrices of the observed radial profiles (within $5R_{500}$), at each frequency, are shown as inserts.}
        \label{fig:Comp_sep}
    \end{figure*}

    \begin{figure*}
    	\includegraphics[width=2\columnwidth]{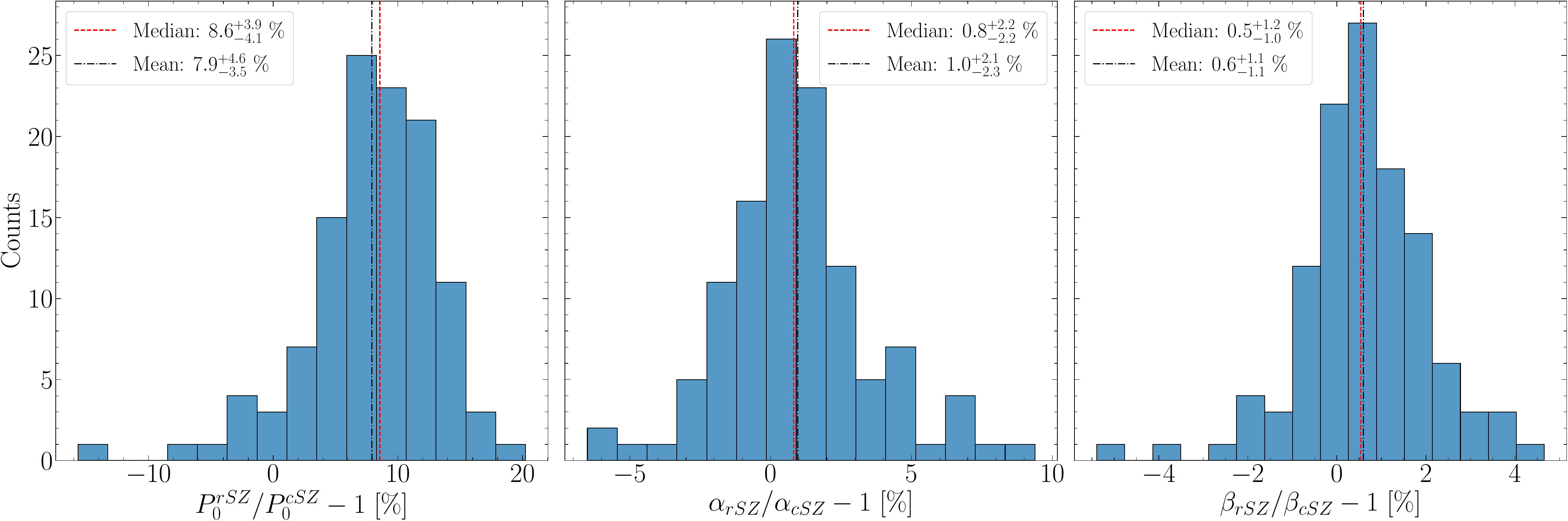}
        \caption{Fractional differences in the pressure profile parameters (from left to right: $P_0$, $\alpha$, $\beta$) between the rSZ and cSZ models. Median (red dashed) and mean (black dot-dashed) are shown with vertical lines and in the legends with $16$th–$84$th percentiles from the reference values.}
        \label{fig:rSZcl_Ppar}
    \end{figure*}
    
    The results of the adopted component separation (up to $12R_{500}$) are summarised in the stacked mean radial profiles shown in Fig.~\ref{fig:Comp_sep}. Specifically, for each HFI channel, we show the mean SZ signal (black crosses) obtained by averaging the observed profiles (Eq.~\ref{eq:I_SZ}) over the 112 clusters with 15 bins within $5R_{500}$ (excluding the remaining four clusters with only 10 bins for consistent stacking; see Sect.~\ref{ssec:mm background}). The CMB (green curves) and GTD (solid red, after the CTD correction with dashed) stacked templates are shown in Fig.~\ref{fig:Comp_sep}, together with the rSZ (solid blue) and cSZ (dashed blue) models. The spreads between the $16$th and $84$th percentiles of the individual profiles are shown as shaded regions. Beyond $\sim4R_{500}$ (but in particular beyond $7R_{500}$, where we fit the CMB and GTD components), we find no significant residuals in the data. Thus, our parametric component separation performs reliably in the outer regions dominated by foregrounds and backgrounds. As in \citetalias{Bourdin2017}, we observe an overall non-zero CTD correction. However, in individual cluster fit, the CTD is difficult to constrain, generally with only upper bounds. 
    
    For the SZ effect, the relativistic model describes the stacked data toward cluster centres better than the classical template, with a reduction in the stacked residuals of about $18\%$ within $3R_{500}$ (and considering all HFI channels). The results are particularly evident in the $353 \,{\rm GHz}$ and $545 \,{\rm GHz}$ channels, which are closer to the maximum of the SZ effect and where relativistic corrections are expected to be more relevant (see e.g. App.~\ref{app:rSZ}). We note that the residuals are computed by comparing the stacked data with the average of individually fitted SZ models, rather than fitting the stacked profiles themselves. In individual fits, the improvement is less evident (as expected for \textit{Planck} data), with a median $\chi^2$ reduction of about $0.4\%$.
    
\begin{figure}
   	\includegraphics[width=\columnwidth]{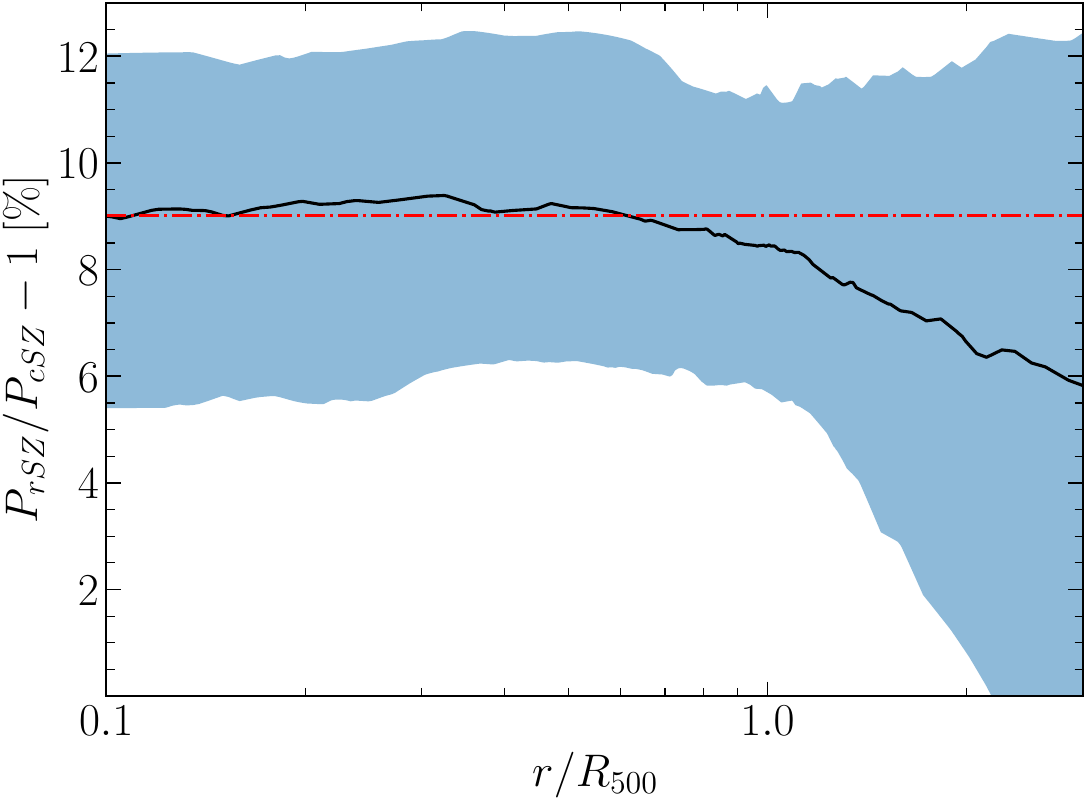}
        \caption{Median fractional change in the pressure profiles between rSZ and cSZ models. The median at $0.1R_{500}$ is shown as a visual reference with the red dot-dashed line, while the blue interval encompasses the $16$th–$84$th percentile range.}
        \label{fig:rSZcl_Prof}
    \end{figure}

    These differences observed in the SZ signal have an impact on the modelled pressure profiles. In Fig.~\ref{fig:rSZcl_Ppar}, we show the relative differences in the pressure parameters. In general, slopes are less affected by rSZ corrections ($\beta_{rSZ}/\beta_{cSZ}-1 \sim 0.5\%$, $\alpha_{rSZ}/\alpha_{cSZ}-1 \sim 0.8\%$) than $P_0$ (about $9\%$). This results in an increase in the pressure profiles, as shown in Fig.~\ref{fig:rSZcl_Prof}. From $0.1$ to $3R_{500}$, the median fractional change spans from $9.0^{+3.0}_{-3.6}\%$ to $5.8^{+6.6}_{-8.8}\%$ (with $16$th–$84$th percentile scatter), showing an almost flat behaviour within $R_{500}$. 
    This trend reflects the joint rSZ and X-ray setup (inclusion of a temperature profile in the SZ spectra, fixed internal slope and $c_{500}$, constraining the \textit{Planck} unresolved inner region with the X-ray data), as well as the residual correlations between $P_0$, $\alpha$, and $\beta$.
    An example of the marginalised posteriors for the pressure profile parameters is shown in Appendix~\ref{app:rSZ} (Fig.~\ref{fig:posteriors}). In general, the change in pressures is comparable to that estimated by \citet{Lee2020} considering clusters at $5 \,{\rm keV}$ (the average temperature of CHEX-MATE clusters is about $6.9 \,\rm keV$).

\subsection{Cluster posteriors and global $\eta_T$ distribution} \label{ssec:eta_results}

    \begin{figure}
    	\includegraphics[width=\columnwidth]{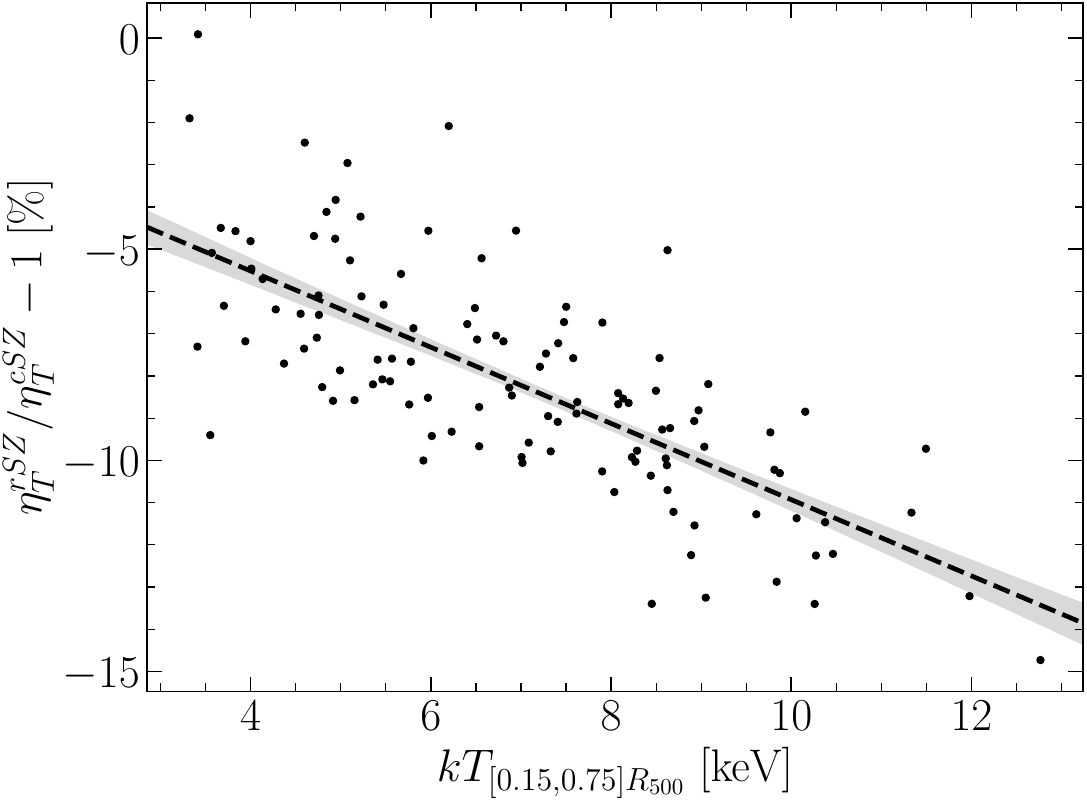}
        \caption{Temperature effect of the relativistic correction on $\eta_T$. Uncertainties on $\eta_T$ values are not shown in the figure, but are at a $14\%$ level. Dashed line (with $1\sigma$ confidence interval) shows the trend of the best linear fit to the data.}
        \label{fig:eta_rcSZTrel}
    \end{figure}

    With or without rSZ corrections, the posterior distributions of $\eta_T$ do not show remarkably Gaussian behaviours \citep[with more skewed distributions, as also noted in][]{Wan2021}, with relative uncertainties of the order of $14\%$. Generally, we find that lognormal distributions better describe the shape of the posteriors, in agreement with the definition of $\eta_T$ as a positive quantity, given by the product of independent terms (Eq.~\ref{eq:eta_def}). Moreover, an anti-correlation with $P_0$ is present (see Fig.~\ref{fig:posteriors} for an example). Thus, we observe a decrease in $\eta_T$, of the same order as $P_0$, when relativistic corrections are included (median $\sim8\%$, see Table~\ref{Tab:eta_stat}). The level of the shift is also temperature dependent, as shown in Fig.~\ref{fig:eta_rcSZTrel}. Considering the temperature estimated from the cluster spectra within $[0.15,0.75]R_{500}$, we observe a linear correlation (Pearson r statistic: $\sim-0.7$), with fractional differences from about $-5\%$ to $-14\%$ in the temperature range $3$--$13 \,{\rm keV}$, similar to what was found for Compton $y$ by \citet{Perrott2024}. However, we note that these shifts are of the same order (or lower) as the actual $1\sigma$ confidence interval on $\eta_T$. Therefore, it is difficult to assess the impact of the temperature dependence of the rSZ corrections on the final results for single clusters, or in reverse, to probe the cluster temperature from relativistic SZ data alone \citep[as also noted by][]{Perrott2024}.

    \begin{figure}
    	\includegraphics[width=\columnwidth]{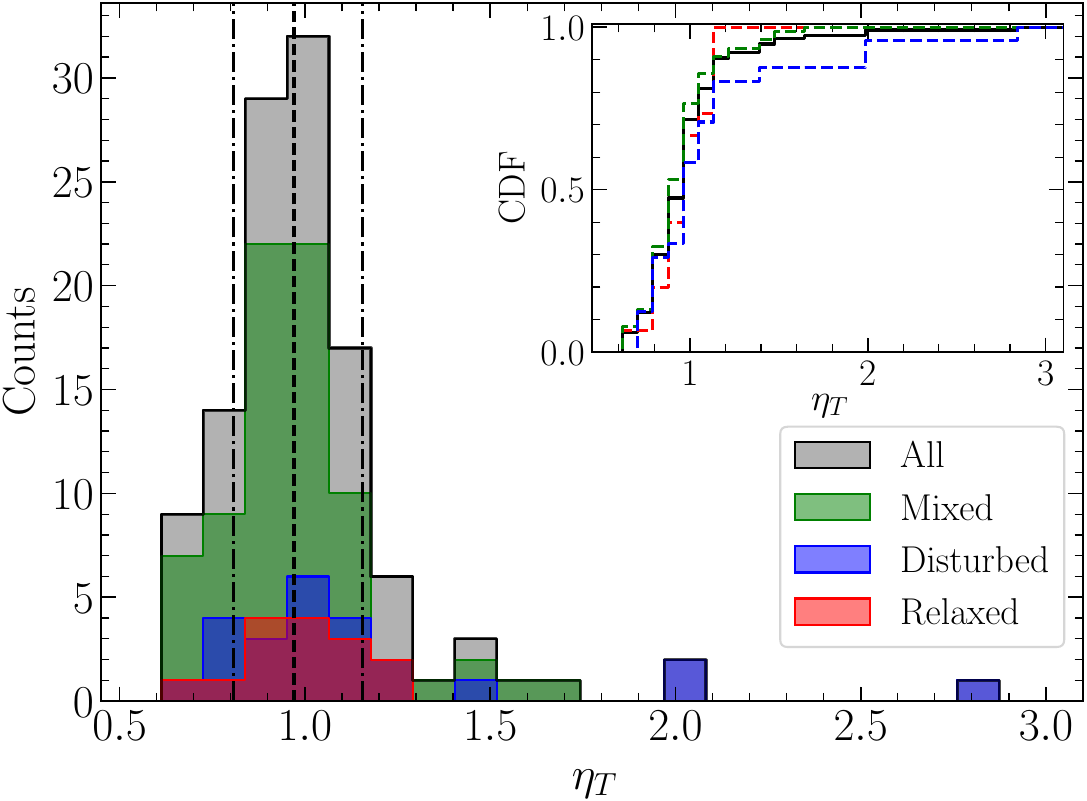}
        \caption{Distributions (and cumulatives in the insert panel) of the $\eta_T$ parameter for the CHEX-MATE sample (black bins). Red, green and blue bins define the distribution of the relaxed, mixed, and disturbed clusters, respectively. Dash dotted and dashed vertical lines show the position of the 16th, 84th percentiles, and median values, respectively.}
        \label{fig:eta_T}
    \end{figure}

    \begin{table}
        \caption{Summary of the $\eta_T$ distribution statistics for the cSZ and rSZ models, Tier-1 and Tier-2 subsamples, the three morphological classes, and the fractional mean and median changes (with $16$th–$84$th percentile variation) given by rSZ corrections and molecular hydrogen fraction in $\rm N_H$.}
        \centering
        \begin{tabular}{@{\hspace{0.1\tabcolsep}} c @{\hspace{0.5\tabcolsep}} c @{\hspace{0.5\tabcolsep}} c @{\hspace{1.\tabcolsep}} c @{\hspace{1.\tabcolsep}} c @{\hspace{1.\tabcolsep}} c @{\hspace{0.5\tabcolsep}} c @{\hspace{0.1\tabcolsep}}}
            \hline\hline
            \multirow{2}{*}{Sample} & \multirow{2}{*}{Mean} & \multirow{2}{*}{Median} & \multirow{2}{*}{$\eta_T^{16th}$} & \multirow{2}{*}{$\eta_T^{84th}$} & $\gamma$ & $\kappa$ \\ 
            & & & & & (skewness) & (kurtosis) \\ 
            \hline 
            cSZ & 1.10 & 1.05 & 0.87 & 1.27 & 2.91 & 12.1 \\ 
            \hline 
            rSZ & 1.01 & 0.97 & 0.81 & 1.15 & 3.20 & 15.1 \\ 
            \hline 
            Tier-1 & 1.01 & 0.96 & 0.81 & 1.12 & 3.49 & 15.1 \\ 
            \hline
            Tier-2 & 1.04 & 0.98 & 0.83 & 1.17 & 3.32 & 14.6 \\ 
            \hline \hline
            Relaxed & 0.99 & 0.99 & 0.86 & 1.17 & -0.42 & -0.3 \\ 
            \hline 
            Mixed & 0.97 & 0.96 & 0.80 & 1.12 & 1.32 & 2.6 \\ 
            \hline
            Disturbed & 1.15 & 0.98 & 0.81 & 1.29 & 2.14 & 4.2 \\ 
            \hline \hline
            \multicolumn{2}{c}{\rule[-1ex]{0pt}{3ex} Fractional change} & \multicolumn{3}{c}{$\eta_T^{rSZ}/\eta_T^{cSZ} - 1 \,[\%]$} & \multicolumn{2}{c}{$\eta_T^{atm}/\eta_T^{tot} - 1 \,[\%]$} \\
            \hline
            \multicolumn{2}{c}{\rule[-1ex]{0pt}{3ex} Mean (Median)} & \multicolumn{3}{c}{$-8.1^{+2.8}_{-2.2}$ ($-8.3^{+3.0}_{-2.0}$)} & \multicolumn{2}{c}{$2.7^{+3.7}_{-3.6}$ ($1.0^{+5.5}_{-1.8}$)} \\
            \hline \hline
        \end{tabular} 
        \label{Tab:eta_stat}
    \end{table}

    In Fig.~\ref{fig:eta_T}, we show the distribution of $\eta_T$ for the 116 CHEX-MATE clusters with the rSZ model, while in Table~\ref{Tab:eta_stat} we summarise the main statistics of the distribution. In general, it has an asymmetric shape with a more prominent tail toward larger values, as pointed out by the skewness and kurtosis ($\gamma$ and $\kappa$ in Table~\ref{Tab:eta_stat}). The bulk of the distribution instead is close to unity (i.e. with the assumption of an ideal ICM distribution, with spherical symmetry, no clumps or bias given in the cosmological framework). Considering the CHEX-MATE subsamples Tier-1 (built to be a local anchor) and Tier-2 (a reference sample for the most massive structures), we do not find strong differences in their $\eta_T$ distributions. From a Kolmogorov–Smirnov test, we have a $p$-value of $0.68$. Thus, we cannot reject the null hypothesis that the two distributions come from the same underlying $\eta_T$ population.

\subsection{Dependence on cluster morphology and physics} \label{sssec:morphology}

    \begin{figure*}
        \includegraphics[width=\textwidth]{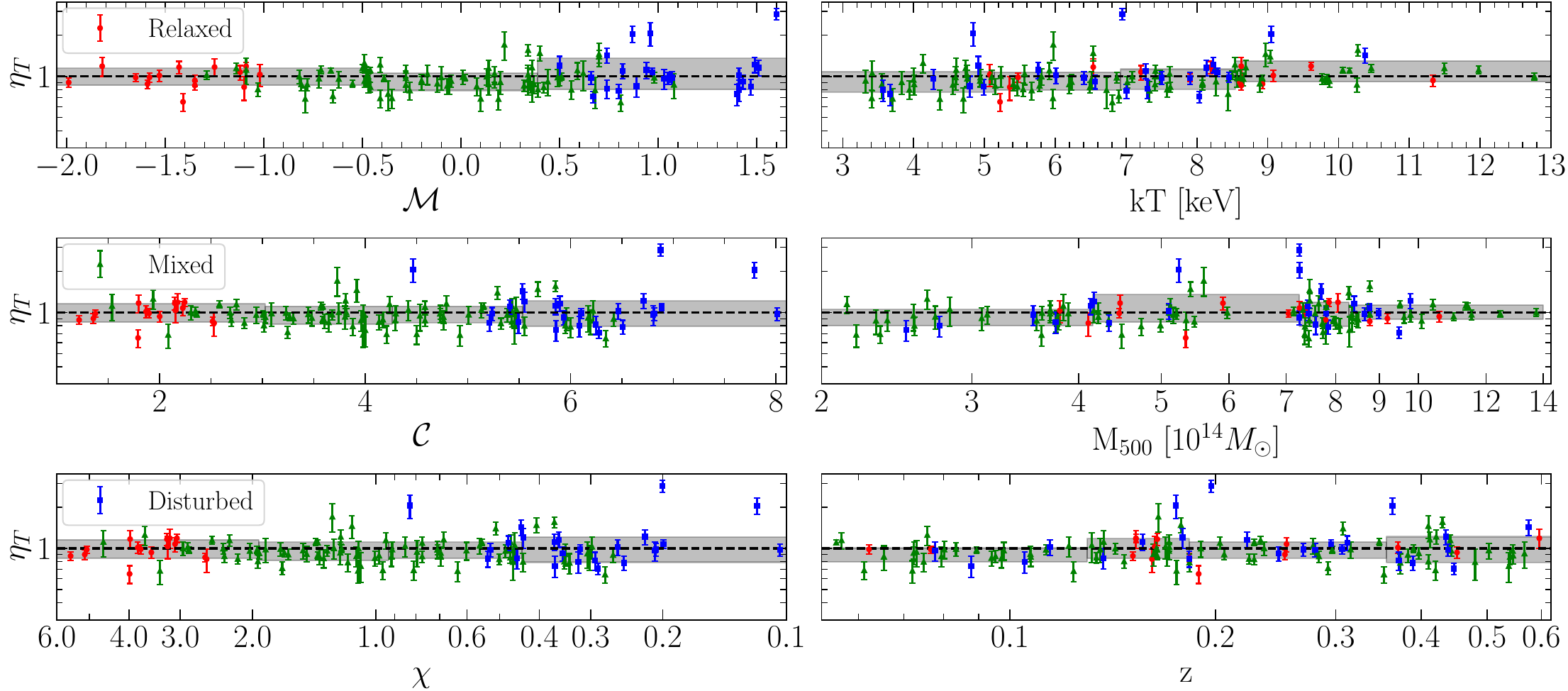}
        \caption{Distribution of $\eta_T$ values (with $1\sigma$ error bars) against cluster morphological indicators $\mathcal{M}$, $\mathcal{C}$, $\chi$ (upper to lower left panels, respectively) and cluster temperatures, masses, and redshifts (right panels). Dynamical classes of \citet{Campitiello2022} are shown as red circles (relaxed), green triangles (mixed), and blue squares (disturbed). Shaded grey envelopes denote the $16$th–$84$th percentile dispersion dividing the sample in quartiles. The x-axis in the $\chi$ panel is reversed, to preserve the same relaxation ordering of $\mathcal{M}$ and $\mathcal{C}$.}
        \label{fig:eta_stat}
    \end{figure*}

    The three outlier clusters with $\eta_T > 2$ (PSZ2G124.20-36.48, PSZ2G143.26+65.24, and PSZ2G218.81+35.51) are all bimodal systems, as shown by their X-ray images (see Appendix~\ref{app:gallery}). For these systems, assumptions of spherical symmetry and smooth thermodynamical profiles are unlikely to hold, and the limited angular resolution of \textit{Planck} prevents a clear separation of the structures in the SZ data. These systematics are effectively captured by $\eta_T$ as discrepancies between the X-ray and SZ profiles. Motivated by this, we examine potential correlations between $\eta_T$ and cluster properties such as morphology, redshift, mass, and temperature. For the CHEX-MATE project, \citet{Campitiello2022} divided the clusters into three dynamical classes: relaxed, mixed, and disturbed. In Fig.~\ref{fig:eta_T} and Table~\ref{Tab:eta_stat}, we show the properties of the $\eta_T$ distributions for these subsamples. Disturbed clusters present a broader distribution and are responsible for the scatter in the $\eta_T$ right tail. Mixed systems are the majority in the CHEX-MATE sample and drive overall statistical trends. Finally, the relaxed subsample is closer and less scattered around $\eta_T = 1$ than the other classes. 
    
    Beyond discrete classification, morphology was continuously quantified by \citet{Campitiello2022} with a set of X-ray indicators, collected with a combined parameter $\mathcal{M}$ \citep{Rasia2013, Cialone2018}. Moreover, Benincasa et al. (in prep.) applied for the first time a Zernike Polynomials-based morphological indicator, $\mathcal{C}$ \citep[introduced in][only for SZ images]{Capalbo2020}, on the CHEX-MATE X-ray cluster observations to estimate the relaxation parameter $\chi$ \citep{Haggar2020,DeLuca2021}. In general, the lower $\mathcal{M}$ and $\mathcal{C}$ are, the more relaxed the cluster is (and the opposite for $\chi$), as shown in the left panels of Fig.~\ref{fig:eta_stat}. No clear trends are present between $\eta_T$ and these indicators. However, more relaxed clusters (leftmost data points) show less scatter, and in some cases smaller uncertainties, than disturbed ones (rightmost points). This trend is also reflected in the higher skewness and kurtosis (Table~\ref{Tab:eta_stat}), as well as in the variation of the $16$th–$84$th percentiles of the sample quartiles (grey intervals in Fig.~\ref{fig:eta_stat}). We note, though, that the CHEX-MATE sample does not collect many relaxed and disturbed systems. Thus, for a more detailed study of this trend on $\eta_T$ (and its impact on cosmological results) could be obtained from hydrodynamical simulations. Regarding any correlation between $\eta_T$ and the mass, temperature, or redshift of the clusters (right panels of Fig.~\ref{fig:eta_stat}), we find weak or no significant trends. In particular, we have Spearman correlation coefficients of about $0.14$ for $M_{500}$ ($p\text{-value} = 0.12$) and $0.31$ for temperature ($p = 6.1 \cdot 10^{-4}$).

\subsection{Other source of systematics} \label{ssec:eta_atom}

    The \textit{Planck} SZ data, although highly sensitive, lack the resolution to probe the inner regions of CHEX-MATE clusters. Thus, we jointly fit millimetre and X-ray data, fixing some SZ pressure parameters (Sect.~\ref{ssec:joint-fit}). Any impact of these fixed parameters on the estimate of $\eta_T$ is expected to be minor due to the strong constraints from the X-ray profiles, although it cannot be excluded for highly disturbed systems with significant substructures (e.g. the bimodal clusters with $\eta_T > 2$). Future improvements can be achieved by including weak lensing data \citep[e.g. as done in][]{Chappuis2025, Saxena2025}, ground-based SZ observations at higher resolutions such as SPT and ACT \citep[for CHEX-MATE analyses, see][]{Oppizzi2022, Gavidia2025}, or with Bolocam, MUSTANG, and NIKA2 \citep{Romero2015, Adam2016, Ruppin2019}.

    \begin{figure*}
    	\includegraphics[width=\columnwidth]{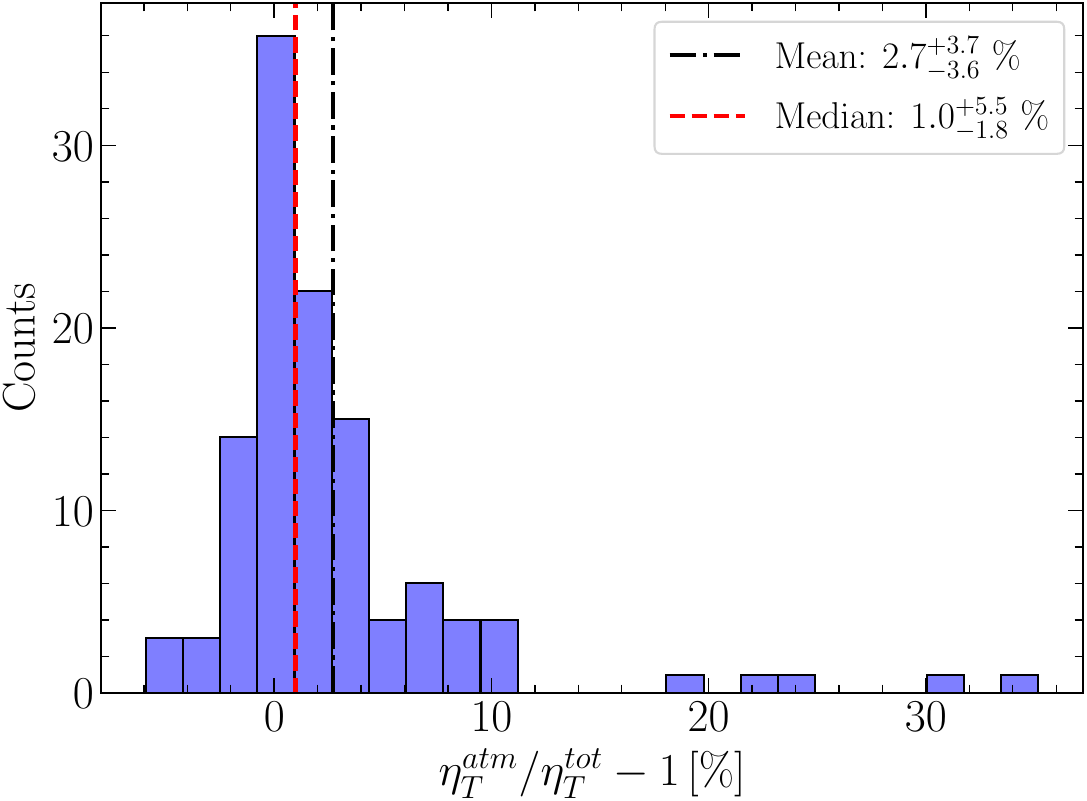}
    	\includegraphics[width=\columnwidth]{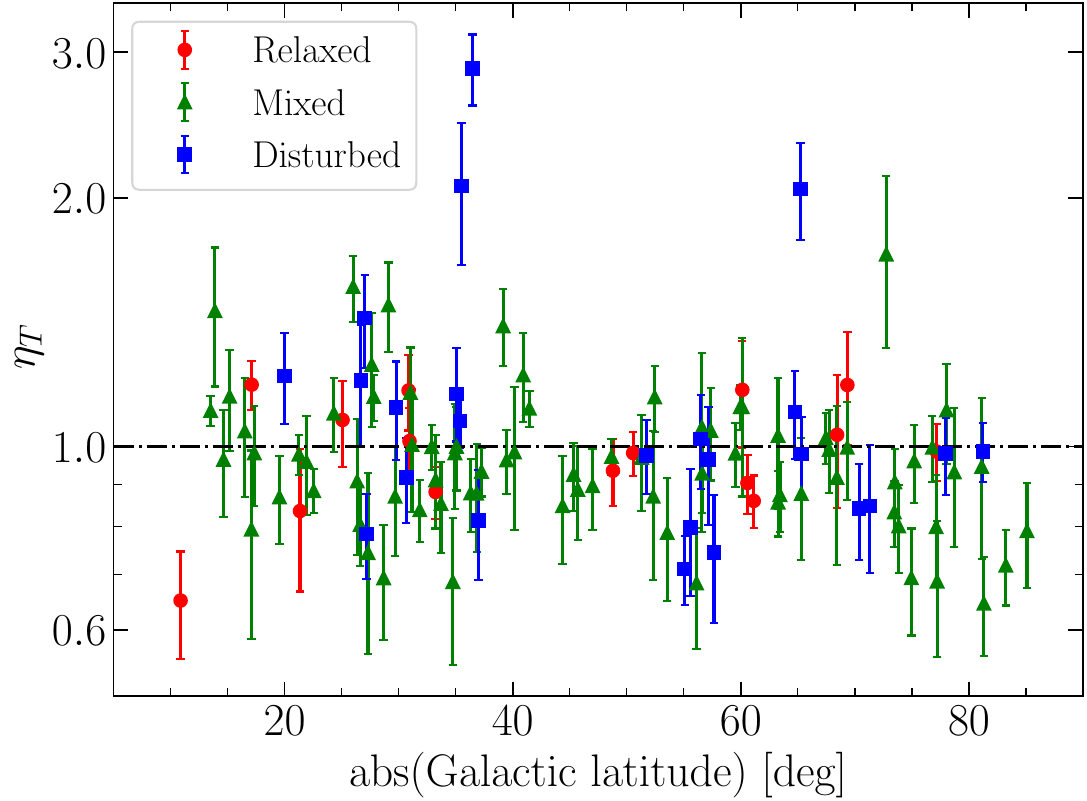}
        \caption{(\textit{left panel}): Fractional change in $\eta_T$ when using the atomic column density in the X-ray analysis. Vertical lines show the median (red) and mean (black) values. (\textit{right panel}): Correlation of $\eta_T$ with the cluster angular distance from the Galactic plane. Dot-dashed line show the $\eta_T = 1$ line as a visual reference. Relaxed, mixed, and disturbed clusters are shown as red circles, green triangles, and blue squares, respectively.}
        \label{fig:eta_T_HI4PI}
    \end{figure*}

    As detailed in \citet{Bourdin2023}, the fraction of molecular hydrogen in the column density must be considered toward the CHEX-MATE clusters for accurate estimates of X-ray spectra, especially for clusters closer to the galactic plane. To assess the impact of the molecular fraction, we repeat the analysis considering only the atomic component. In general, a lower $\rm N_H$ yields higher X-ray spectroscopic temperatures and slightly lower emission measures. We then expect (from Eq.~\ref{eq:eta_T}) an increase in $\eta_T$, since temperature estimates are more sensitive to absorption model variations. This effect is shown in the left panel of Fig.~\ref{fig:eta_T_HI4PI} and summarised in Table~\ref{Tab:eta_stat}. A change is observed, with a median (mean) variation of about $1\%$ ($2.7\%$), but reaching values up to $34\%$ for some clusters. This small variation reflects the selection of CHEX-MATE clusters. Most of them (about $80\%$ of the total) are in regions where the molecular content is negligible, or its contribution is lower than $10\%$ on average, as shown by \citet{Bourdin2023}. Moreover, residual contamination from Galactic dust emission can introduce systematics in our analysis. As in \citet{Wan2021}, we tested the robustness of our results by checking for $\eta_T$ correlations with the distance from the Galactic plane, as shown in the right panel of Fig.~\ref{fig:eta_T_HI4PI}. If the contamination were significant, clusters closer to the Galactic plane would be more affected. However, no significant trend is observed.
    
    As a final consideration, we note that current X-ray telescopes, such as Chandra, XMM-\textit{Newton}, and even eROSITA exhibit systematic discrepancies in the spectroscopic temperatures \citep[e.g. see][]{Nevalainen2010, Schellenberger2015, Migkas2024}. Moreover, these discrepancies are temperature dependent, with larger deviations at higher temperatures. Thus, $\eta_T$ measured with different telescopes are expected to differ, with XMM-\textit{Newton} generally yielding lower values compared to Chandra, as also shown in \citetalias{Bourdin2017}. The mean $\eta_T$ value of $1.01 \pm 0.03$ (standard error) for the CHEX-MATE clusters denotes good agreement between XMM-\textit{Newton} temperatures and millimetre estimates (and, in particular, for the relaxed subsample where biases are expected to be smaller). However, a detailed description of the cross-calibration issues and their impact on the values of $\eta_T$ is beyond the scope of this work, based on a follow-up of \textit{Planck}-selected clusters observed with XMM-\textit{Newton}.

\section{Discussion and Conclusions} \label{sec:Conclusions}

    In this work, we present a new study of systematic mismatches in temperature estimates derived from a joint analysis of XMM-\textit{Newton} and \textit{Planck}-HFI observations. Discrepancies between X-ray and millimetre estimates are expected, as the usual simplifying assumptions for the ICM distribution (e.g. spherical symmetry, and regular profiles) only approximate the gas distribution. Thus, these mismatches can be used to improve cosmological constraints, our knowledge of the cluster structure, or to cross-calibrate temperature measurements (\citetalias{Kozmanyan2019}; \citealp{Ettori2020, Wan2021}). Compared to previous studies, this work extends the analysis to a larger sample of 116 galaxy clusters from the PSZ2 catalogue, selected for the CHEX-MATE project to study the local cluster population (Tier-1 subsample) and (Tier-2) massive systems \citep{CHEX-MATE}. In particular, we improve the parametric methodology of \citetalias{Bourdin2017} including a more detailed X-ray characterisation of soft proton contamination, the molecular fraction in the hydrogen column density \citep[as estimated in][]{Bourdin2023}, and relativistic corrections in the SZ modelling, based on X-ray temperatures. Our findings can be summarised as follows.
    
    When relativistic corrections are included, the residuals within $3R_{500}$ in the mean stacked profiles (Fig.~\ref{fig:Comp_sep}) are reduced by approximately $18\%$. In addition, the pressure profiles (Fig.~\ref{fig:rSZcl_Prof}) show a noteworthy fractional increase of about $8\%$, with a gradient from $0.1 R_{500}$ (median value and $16$th–$84$th dispersion: $9.0^{+3.0}_{-3.6}\%$) to $3R_{500}$ ($5.8^{+6.6}_{-8.8}\%$). Thus, given the anti-correlation between the normalisation of the pressure profiles and $\eta_T$, we find a similar change for $\eta_T$, on average of $-8.1^{+2.8}_{-2.2}\%$ (median: $-8.3\%$). The shift depends on the cluster temperature, ranging from about $-5\%$ to $-14\%$ over the $3$--$13 \, {\rm keV}$ interval. These changes in $\eta_T$ (if we consider these as systematic biases in Compton $y$) are similar to those found in previous studies, such as in \citet{Lee2020} and \citet{Perrott2024}. However, in individual cluster fits, the relativistic model is only marginally favoured, and the relative uncertainties in $\eta_T$ (about $14\%$) are too large to constrain temperatures from relativistic SZ spectra. 
    \\
    The inclusion of the molecular content in the X-ray absorption model is crucial for accurate temperature estimates \citep{Bourdin2023}. If not considered, we observe for the CHEX-MATE sample a mean shift on $\eta_T$ (and thus on X-ray temperatures) of $2.7^{+3.7}_{-3.6}\%$ (median: $1\%$, but it could be up to $34\%$ for some cluster).
    
    We find little or no correlations between $\eta_T$ and cluster masses, redshifts, and temperatures (Fig.~\ref{fig:eta_stat}). Moreover, we do not observe relevant differences between the $\eta_T$ distributions of Tier-1 and Tier-2 subsamples. Considering the CHEX-MATE morphology studied in \citet{Campitiello2022} and Benincasa et al. (in prep.), the dynamical state of the clusters mainly affects the scatter of $\eta_T$ values (Table~\ref{Tab:eta_stat} and Fig.~\ref{fig:eta_T}). The CHEX-MATE $\eta_T$ distribution has a mean of $1.01\pm0.03$ (median $0.97$) and is positively skewed, with relaxed clusters having values closer to unity and less scattered than mixed and disturbed systems, prevalent in the right tail. Thus, samples with less relaxed systems are prone to present more outliers which, if unaccounted for, can bias subsequent analyses such as cosmological constraints \citepalias{Kozmanyan2019}. Based on these results, we will use a large sample of simulated clusters to extend the investigation of the relation between $\eta_T$ and the cluster dynamical state. The information derived from the simulations, in combination with the results of this work, will then be used to derive new cosmological constraints on $H_0$ extending the methodology of \citetalias{Kozmanyan2019}.

\begin{acknowledgements}
    
    We thank the anonymous referee for useful comments. HB, FDL, SE, PM, MR, MS acknowledge the financial contribution from the contracts Prin-MUR 2022 supported by Next Generation EU (M4.C2.1.1, n.20227RNLY3 \textit{The concordance cosmological model: stress-tests with galaxy clusters}), MDP from PRIN-MUR grant 20228B938N {\it"Mass and selection biases of galaxy clusters: a multi-probe approach"} funded by the European Union Next generation EU, Mission 4 Component 1 CUP C53D2300092 0006, MS from contract INAF mainstream project 1.05.01.86.10 and INAF Theory Grant 2023: Gravitational lensing detection of matter distribution at galaxy cluster boundaries and beyond (1.05.23.06.17), LL from the INAF grant 1.05.12.04.01. HB, FDL, PM also acknowledge the support by the Fondazione ICSC, Spoke 3 Astrophysics and Cosmos Observations, National Recovery and Resilience Plan (Piano Nazionale di Ripresa e Resilienza, PNRR) Project ID CN\_00000013 "Italian Research Center on High-Performance Computing, Big Data and Quantum Computing" funded by MUR Missione 4 Componente 2 Investimento 1.4: Potenziamento strutture di ricerca e creazione di "campioni nazionali di R\&S (M4C2-19 )" - Next Generation EU (NGEU), and by INFN through the InDark initiative, M.M.E. and E.P. the support of the French Agence Nationale de la Recherche (ANR), under grant ANR-22-CE31-0010, M.G. from the ERC Consolidator Grant \textit{BlackHoleWeather} (101086804), DE from the Swiss National Science Foundation (SNSF) under grant agreement 200021\_212576, BJM from Science and Technology Facilities Council grants ST/V000454/1 and ST/Y002008/1. H.S and J.S. were supported by NASA Astrophysics Data Analysis Program (ADAP) Grant 80NSSC21K1571. AF acknowledges the project "Strengthening the Italian Leadership in ELT and SKA (STILES)", proposal nr. IR0000034, admitted and eligible for funding from the funds referred to in the D.D. prot. no. 245 of August 10, 2022 and D.D. 326 of August 30, 2022, funded under the program "Next Generation EU" of the European Union, “Piano Nazionale di Ripresa e Resilienza” (PNRR) of the Italian Ministry of University and Research (MUR), “Fund for the creation of an integrated system of research and innovation infrastructures”, Action 3.1.1 "Creation of new IR or strengthening of existing IR involved in the Horizon Europe Scientific Excellence objectives and the establishment of networks”. This research was supported by the International Space Science Institute (ISSI) in Bern, through ISSI International Team project \#565 ({\it Multi-Wavelength Studies of the Culmination of Structure Formation in the Universe}), the Basic Science Research Program through the National Research Foundation of Korea (NRF) funded by the Ministry of Education (2019R1A6A1A10073887), the 2025 KAIST-U.S. Joint Research Collaboration Open Track Project for Early-Career Researchers, supported by the International Office at the Korea Advanced Institute of Science and Technology (KAIST). Work at Argonne National Lab is supported by UChicago Argonne LLC, Operator of Argonne National Laboratory (Argonne). Argonne, a U.S. Department of Energy Office of Science Laboratory, is operated under contract no. DE-AC02-06CH11357. FDL, HB, and PM also thank the INFN Roma2 IT group for their invaluable support, particularly in the aftermath of the recent fire at their facilities. This work made use of {\sc IDL} Astronomy Users’s Library \citep{IDLastro}, HEASoft \citep{Heasoft2014}, HEALPix \citep{healpix, healpix2} and diverse {\sc python} packages: {\sc numpy} \citep{numpy}, {\sc scipy} \citep{SciPy}, {\sc matplotlib} \citep{matplot}, {\sc seaborn} \citep{seaborn}, {\sc pandas} \citep{pandas, pandas2}, {\sc astropy} \citep{astropy,astropy2,astropy3}.

\end{acknowledgements}

%
\bibliographystyle{aa}
\bibliography{biblio}

\begin{appendix}

\section{Comparison with the standard CHEX-MATE pipeline} \label{app:T_comparison}
    
    \begin{figure*}
    	\includegraphics[width=0.915\columnwidth]{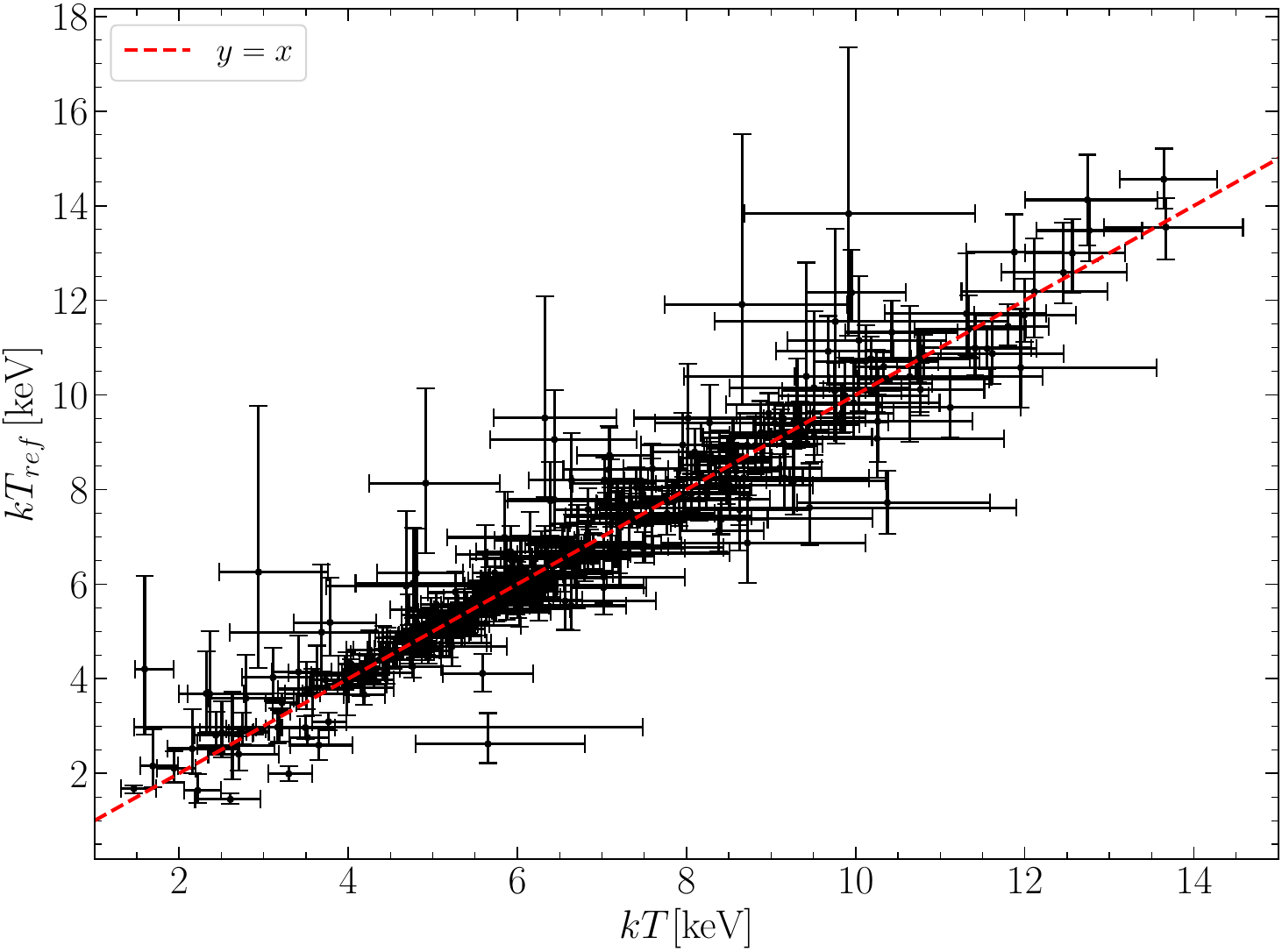}
        \hspace{1cm}
    	\includegraphics[width=\columnwidth]{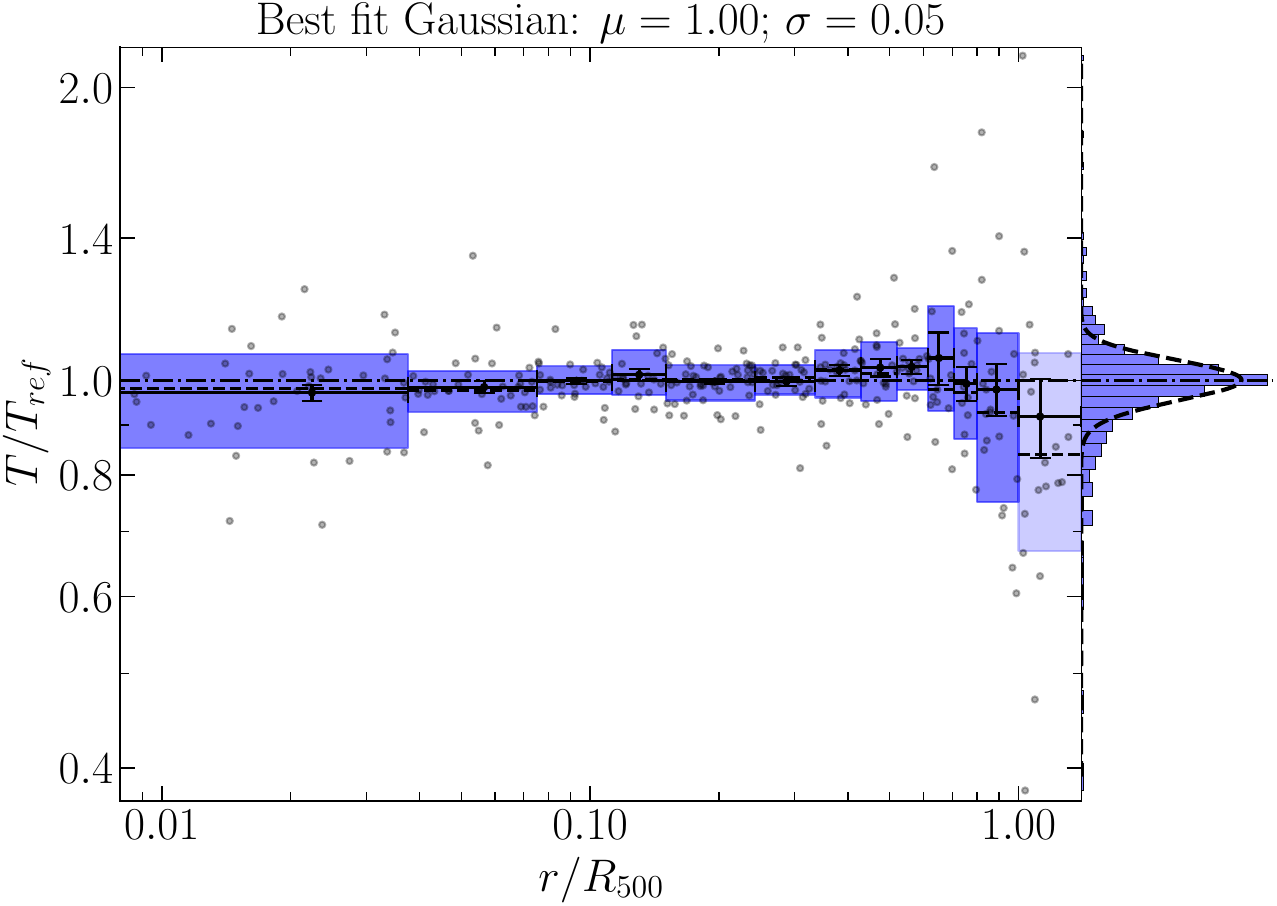}
        \caption{(\textit{Left panel}): Comparison between the CHEX-MATE pipeline temperature ($T_{ref}$) and the results of this work, with $1\sigma$ error bars and the equality line in red dashed. (\textit{Right panel}): Ratios between the temperature estimates (grey dots) as a function of the radial distance from the cluster centres. The overall distribution is shown in the right marginal plot, with a Gaussian fit. Dark blue areas show the $16$th–$84$th percentile range in the radial binning of this work. The bin mean (with its standard uncertainty) and median values are shown with black dots (with error bars) and dashed segments. As a visual reference, the $y=1$ line is shown with a dot-dashed line.}
        \label{fig:T_comp}
    \end{figure*}
    
    The results shown in this paper are based on an independent analysis of the CHEX-MATE galaxy cluster sample. This choice was made to extend the already developed cosmological pipeline of \citetalias{Kozmanyan2019} to a larger sample of clusters (almost a factor two) and, at the same time, to include more sources of systematics in the model. Differences from the CHEX-MATE standard pipeline mainly concern the X-ray background modelling, including flare filtering and the calibration of the SP component. In \citet{Rossetti2024}, the authors conducted an accurate estimate of the SP and QPB components for the background, considering the correlations between the count rates inside and outside the regions subtended by the FOV for the EPIC cameras, similarly to the method of \citet{DeLuca2004}, \citet{Leccardi2008}, and \citet{Ghirardini2018}. In particular, the scaling of the SP component with the IN-OUT method from \citet{Ghirardini2018} is extended to spectral analysis, and uncertainties in the background model are propagated to cluster spectra (e.g. for temperature and metallicity profiles) considering an MCMC fit. For this reason, we test the consistency between our results and the CHEX-MATE outcomes for 28 clusters of the DR1 subsample presented in \citet{Rossetti2024}. For this analysis, we adopt the same radial binning, cluster modelling (e.g. $\rm N_H$, abundance table, cross-sections) and spectral band ($0.5$--$12 \,\rm keV$) used for temperature estimates. 
    
    A comparison of the two pipelines is shown in Fig.~\ref{fig:T_comp}. In the left panel, we show the direct comparison (with $1\sigma$ uncertainties) of the temperature results of this work ($kT$, x-axis) and of the CHEX-MATE pipeline ($kT_{ref}$, y-axis), while the right panel presents the radial distribution of their ratio. In general, the temperatures are close to the equality line, and the distribution of their ratio is close to being symmetric and centred on $1$. However, the distribution has a non-zero excess kurtosis, with heavier tails than the fitted Gaussian. From the radial profile, the agreement between the two pipelines is stronger toward the cluster centres and at intermediate radii, while the outliers occur mostly in the cluster outskirts (beyond $R_{500}$), where the S/N is lower and the measures are more challenging. We note, however, that our last bin has edges between $0.8-1R_{500}$. Therefore, the two pipelines yield consistent temperature estimates in the radial range explored in this work (shown in the figure with the dark blue bins).

\section{Relativistic correction for thermal SZ effect} \label{app:rSZ}

\begin{figure}
        \centering
    	\includegraphics[width=\columnwidth]{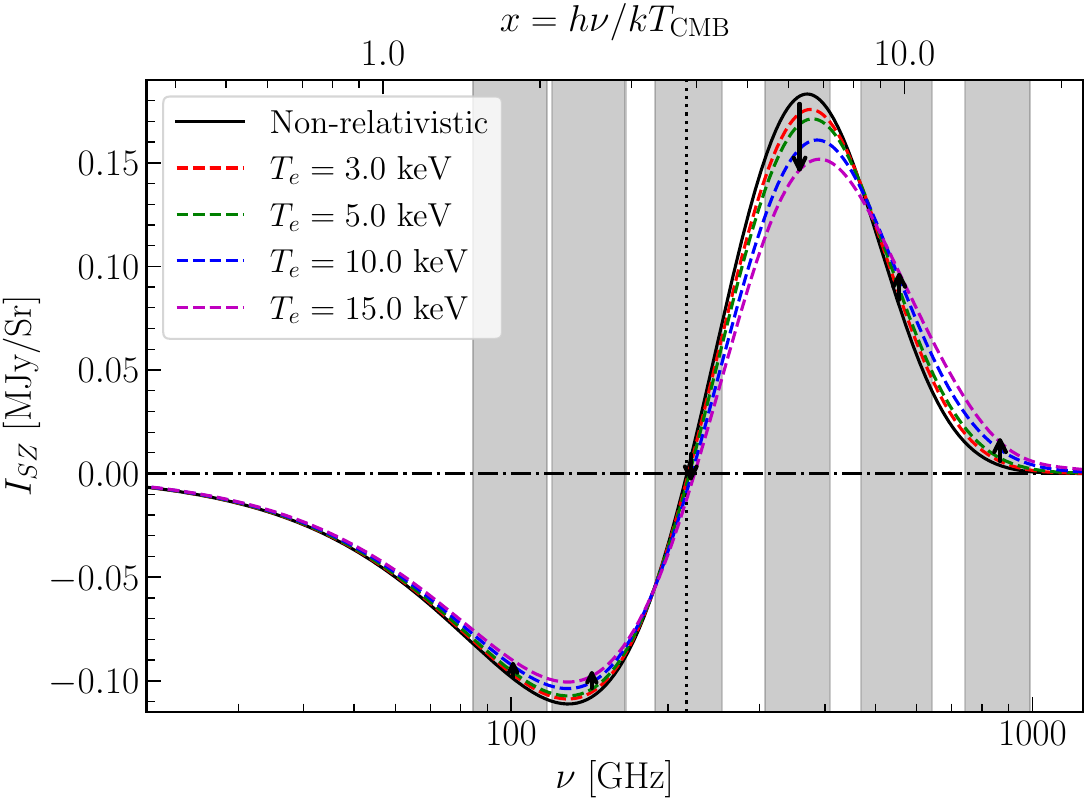}
        \caption[]{Comparison between the non-relativistic (solid black line) and relativistic (dashed) tSZ spectra for fixed $y=10^{-4}$. The dotted line marks the non-relativistic crossover frequency. Grey intervals show the HFI channels, while arrows mark the direction of the $y$ biases induced by non-relativistic modelling.}      
        \label{fig:rSZ_example}
    \end{figure}

    In the classical approximation, the tSZ amplitude (i.e. the $y$; Eqs.~\ref{eq:I_y},~\ref{eq:y}) depends on the electron pressure (and thus on $T_e$), while the spectrum is independent of the electron temperature. As shown in Fig.~\ref{fig:rSZ_example}, the classical spectrum exhibits a decrement of intensities at lower frequencies and an increment at higher frequencies (due to inverse Compton scattering of CMB photons), with a crossover frequency at around $217 \,{\rm GHz}$. When relativistic electrons are included, the spectra become temperature dependent and more asymmetric. For increasing temperatures we generally have a reduction of the peak-to-trough amplitude and a redistribution of the signal towards the Wien tail, with crossover and peak positions shifting towards higher frequencies. As a consequence, classical modelling can lead to biased estimates of the $y$ parameter, with the sign and magnitude depending on the observed frequencies (and the instrumental response). As shown in Fig.~\ref{fig:rSZ_example} for the HFI bands, the non-relativistic formulation generally overestimates the signal at frequencies below the crossover and near the peak. At higher frequencies ($\gtrsim 480 \,{\rm GHz}$), we have instead an underestimation. Therefore, for a proper characterisation of the cluster signal (and of the derived thermodynamical information) a proper relativistic treatment is necessary.
    
\begin{figure*}
        \centering
        \includegraphics[width=\textwidth]{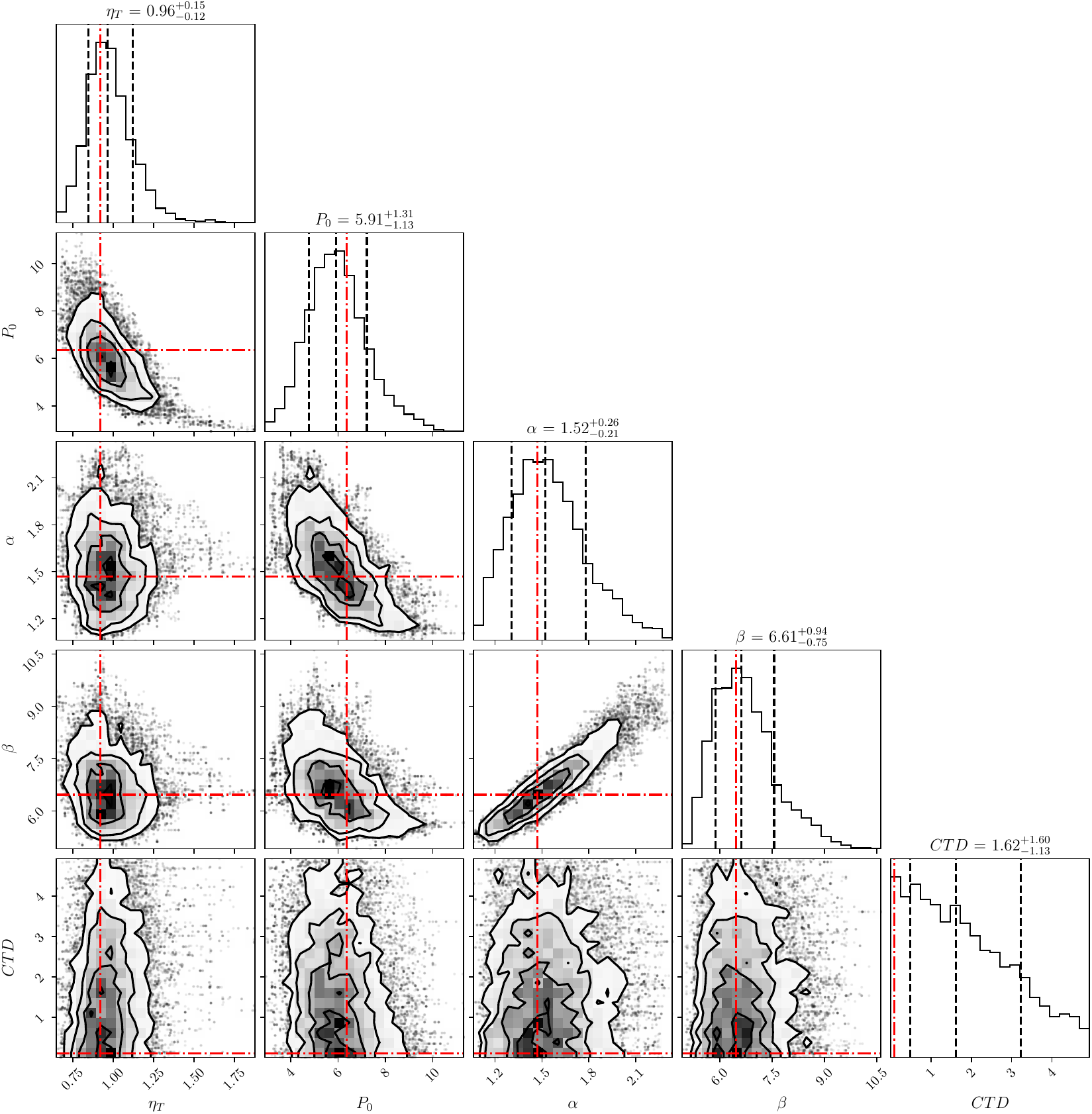}
        \caption{Posterior distributions of the pressure profile parameters for cluster PSZ2G325.70+17.34. Values in captions and dashed black lines report the median, 16th, and 84th percentiles (with plus minus intervals from the median values). Red dot-dashed lines show the position of the maximum likelihood value estimates.}
        \label{fig:posteriors}
    \end{figure*}

    Assuming an isothermal distribution for the gas, the tSZ signal, line-of-sight integrated and corrected for relativistic effect, can be written as:
    \begin{equation}
        I_{SZ}(x, \theta_e) = \frac{x^3}{e^x-1} \frac{\Delta n(x, \theta_e) }{n_0} = \frac{x^4 e^x}{(e^x-1)^2} \vec{\mathcal{Y}}(x) \cdot \vec{\Theta_e}(n_e, \theta_e),
        \label{eq:nozawa}
    \end{equation}
    where $\vec{\mathcal{Y}}(x)$ and $\vec{\Theta_e}$ are the vectors with the expansion coefficients of \citet{Nozawa1998, Nozawa2006} and their associated $\theta_e$ powers, respectively:
    \begin{align}
        &\vec{\mathcal{Y}}(x) = \left[ Y_0(x) , Y_1(x) , Y_2(x) , Y_3(x) , Y_4(x) \right], \label{eq:Y(x)}
        \\
        &\vec{\Theta_e}(n_e, \theta_e) = \sigma_T n_e \left[ \theta_e , \theta_e^2 , \theta_e^3 , \theta_e^
        4, \theta_e^5 \right],
        \label{eq:Theta(Te)}
    \end{align}
    up to the fourth order and where the zero order $Y_0(x)$ is the Kompaneets solution. With this notation (assuming an isothermal gas), the SZ signal can still be written separating the spatial ($\vec{\Theta_e}$) and spectral ($ \vec{\mathcal{Y}}$) behaviours. However, the ICM in galaxy clusters is not in an isothermal state and the temperature can vary from cluster cores to the outskirts. To account for line-of-sight temperature variations, we follow a similar approach to \citet{Prokhorov2012} and \citet{Chluba2013}. Considering Eq.~\eqref{eq:nozawa} as the spectrum given for a reference temperature $\bar{\theta}_e$ along the line of sight, we can introduce corrections for the temperature variation with a second expansion of $I_{SZ}$ around $\bar{\theta}_e$:
    \begin{equation}
    \begin{aligned}
        \int F(x, \theta_e) \,dl &= \int F(x, \bar{\theta}_e) \,dl \, +
        \\ 
        &+ \vec{\mathcal{Y}}(x)  \cdot \sum_{k\ge 1} \int \frac{1}{k!} \left. \frac{\partial^k \vec{\Theta_e}}{\partial \theta_e^k} \right|_{\theta_e=\bar{\theta}_e} \left(\theta_e - \bar{\theta}_e \right)^k dl,
    \end{aligned}
    \label{eq:mom expansion}
    \end{equation}
    with $F(x, \theta_e)$ and the derivatives defined, respectively, as:
    \begin{align}
        &F(x, \theta_e)\, dl = \vec{\mathcal{Y}}(x) \cdot \vec{\Theta_e}(n_e, \theta_e) \,dl = \sigma_T n_e \sum_{i=0}^n  \theta_e^{i+1} Y_i(x) \,dl ,
        \label{eq:F(x)}
        \\
        &\left. \frac{1}{k!} \partial^k_{\theta_e} \Theta_e^i (n_e, \theta_e) \right|_{\theta_e=\bar{\theta}_e} =
    \begin{cases}
         \binom{i+1}{k} \sigma_T n_e \bar{\theta_e}^{i+1-k} & \text{if } k \leq i+1,
        \\
        0 & \text{otherwise}.
    \end{cases}
    \label{eq:deriv theta}
    \end{align}
    As noted in \citet{Chluba2013}, the terms $(\theta_e - \bar{\theta}_e)^k$ in Eq.~\eqref{eq:mom expansion} are equivalent to introduce corrections given by the temperature distribution moments $\mu_k$ about the value $\bar{\theta}_e$ (along the LOS). Thus, using standardised central moments about the mean, it is possible to simplify Eq.~\eqref{eq:mom expansion}. For example, considering the first four central moments and defining the reference temperature as the density-weighted temperature, similarly to \citet{Prokhorov2012} \citep[or $\tau$-weighted, as in Eq.~15 of][]{Chluba2013}, we have:
    \begin{equation}
        \bar{\theta}_e \equiv \frac{\int n_e \theta_e \, dl}{\int n_e \, dl} = \frac{\int \theta_e \, d\tau}{\int \,d\tau},
    \end{equation}
    \begin{align}
    &\int \frac{1}{k!} \left. \partial^k_{\theta_e} \Theta_e^i \right|_{\theta_e=\bar{\theta}_e} \left(\theta_e - \bar{\theta}_e \right)^k dl = \binom{i+1}{k} \bar{\theta}_e^{i-k}  \mu_k \, y(n_e, \theta_e),
    \\
    &\mu_k = \frac{\int n_e \left(\theta_e - \bar{\theta}_e \right)^k dl}{\int n_e dl} =
    \begin{cases}
        0                   & \text{if } k = 1, \\
        \sigma^2_e          & \text{if } k = 2, \\
        \gamma_e \sigma^3_e & \text{if } k = 3, \\
        \kappa_e \sigma^4_e & \text{if } k = 4. \\
    \end{cases}
    \label{eq:Tmom}
    \end{align}
    Therefore, the first central moment is zero by definition, and the second, third, and fourth are given, respectively, by the temperature variance $\sigma^2_e$, skewness $\gamma_e$, and kurtosis $\kappa_e$, all estimated along the LOS Here, $y(n_e, \theta_e)$ is the Compton $y$ parameter, and it is introduced to replicate the general notation used for the tSZ effect (e.g. Eq.~\ref{eq:nozawa}), where it is factorisable as the product of $y$ (which depends on the properties of the gas we are interested in, e.g. the pressure) times a spectrum (with the average gas properties) as in Eq.~\eqref{eq:I_rSZ}. For the calculation of the relativistic corrections described above, we use the temperature and density functional forms proposed by \citet{Vikhlinin2006}, fitted to the X-ray data as described in Sect.~\ref{ssec:X-ray cl sp model}. In Fig.~\ref{fig:posteriors}, we show an example of the outcome of the MCMC fitting of the rSZ signal for the cluster PSZ2G325.70+17.34.

\section{Galaxy cluster gallery} \label{app:gallery}

    Fig.~\ref{fig:Gallery} is a gallery of the 118 galaxy clusters of the CHEX-MATE sample. Every image is produced using a sparse denoising algorithm based on a wavelet reconstruction of the maps \citep{Starck2009, Bourdin2013}. Poisson noise is removed from the images using soft thresholding of B3-spline wavelet transforms, with corrections for the exposure and the vignetting of the XMM-\textit{Newton} EPIC cameras. The maps are also background subtracted, and all the point sources are masked and inpainted. Each map subtends a region of $2R_{500} \times 2R_{500}$, with the X-ray surface brightness extracted in the soft ($0.5, 2.5 \rm keV$) X-ray energy band. The maps are shown in logarithmic scale. With respect to the gallery presented in Fig.~6 of \citet{CHEX-MATE}, Fig.~\ref{fig:Gallery} also collects observations not available at that time and considers a different reconstruction and smoothing of the images \citep[e.g. a wavelet reconstruction here, while a Gaussian smoothing in][]{CHEX-MATE}. The images are ordered according to the combined morphological parameter $\mathcal{M}$ \citep[see Table~A1 in ][]{Campitiello2022}, from the most relaxed to the most disturbed clusters. As discussed in \citet{Campitiello2022}, this sample includes a variety of clusters with different dynamical states. In particular, this sample is not dominated by relaxed systems and presents several objects with disturbed morphologies, such as A115 (PSZ2G124.20-36.48), A1430 (PSZ2G143.26+65.24), or PSZ2G218.81+35.51, all with $\eta_T >2$. The clusters excluded from this work are PSZ2G283.91+73.87, affected by the foreground contamination of the Virgo cluster, and PSZ2G028.63+50.15. Both are shown in the gallery with their names in red.
    
    \begin{figure*}
        \centering
    	\includegraphics[width=0.971\textwidth]{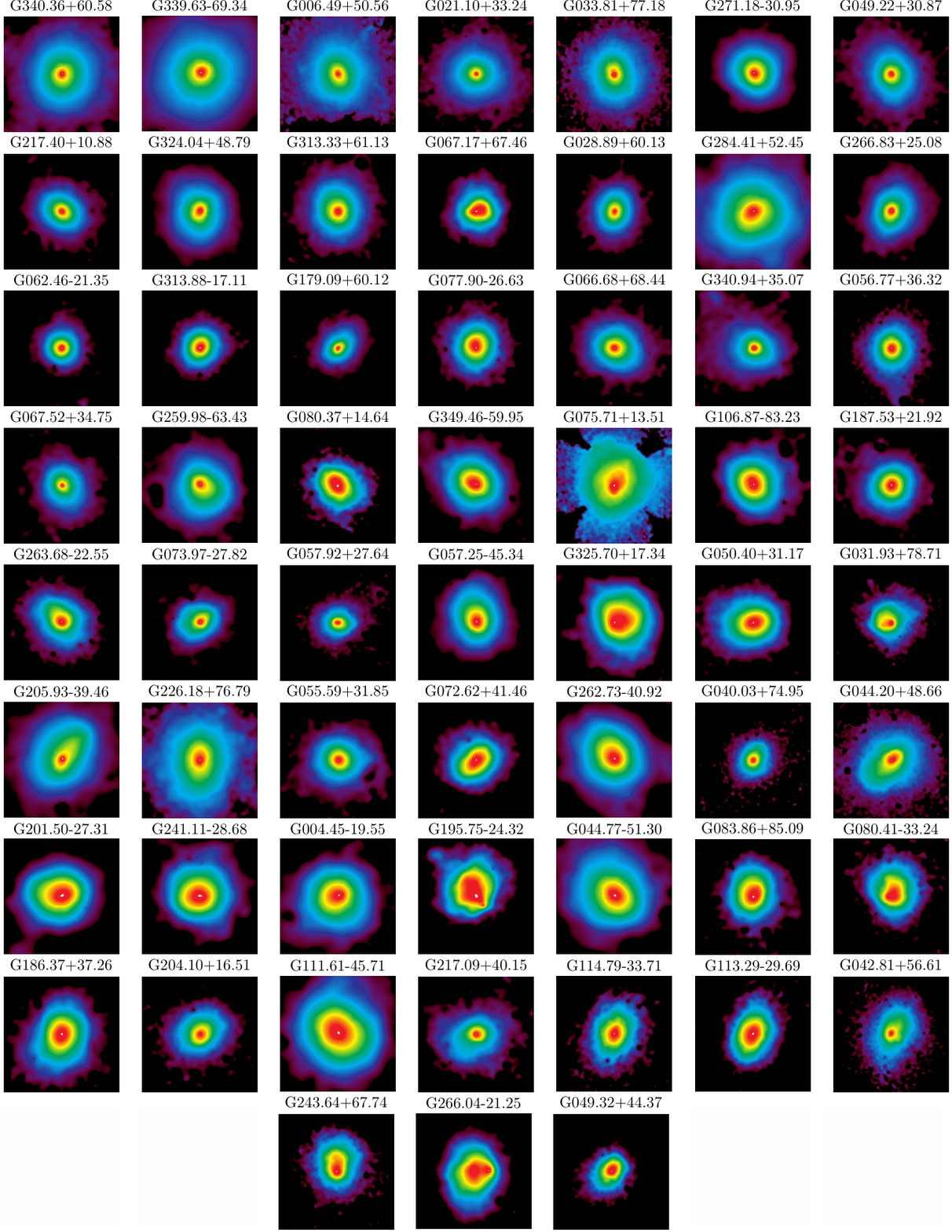}
        \caption{Wavelet-denoised XMM-\textit{Newton} images of the 118 CHEX-MATE galaxy clusters, background subtracted and corrected for vignetting and exposure. Surface brightness in the energy range $[0.5,2.5]\,\rm keV$ is shown with a logarithmic scale from $0.1$ counts per pixel to the image maximum, with point sources removed and inpainted. The maps are centred on the X-ray peak and covers an area of $2R_{500} \times 2R_{500}$.}
        \label{fig:Gallery}
    \end{figure*}
    
    \setcounter{figure}{0}
    \begin{figure*}
        \centering
    	\includegraphics[width=0.971\textwidth]{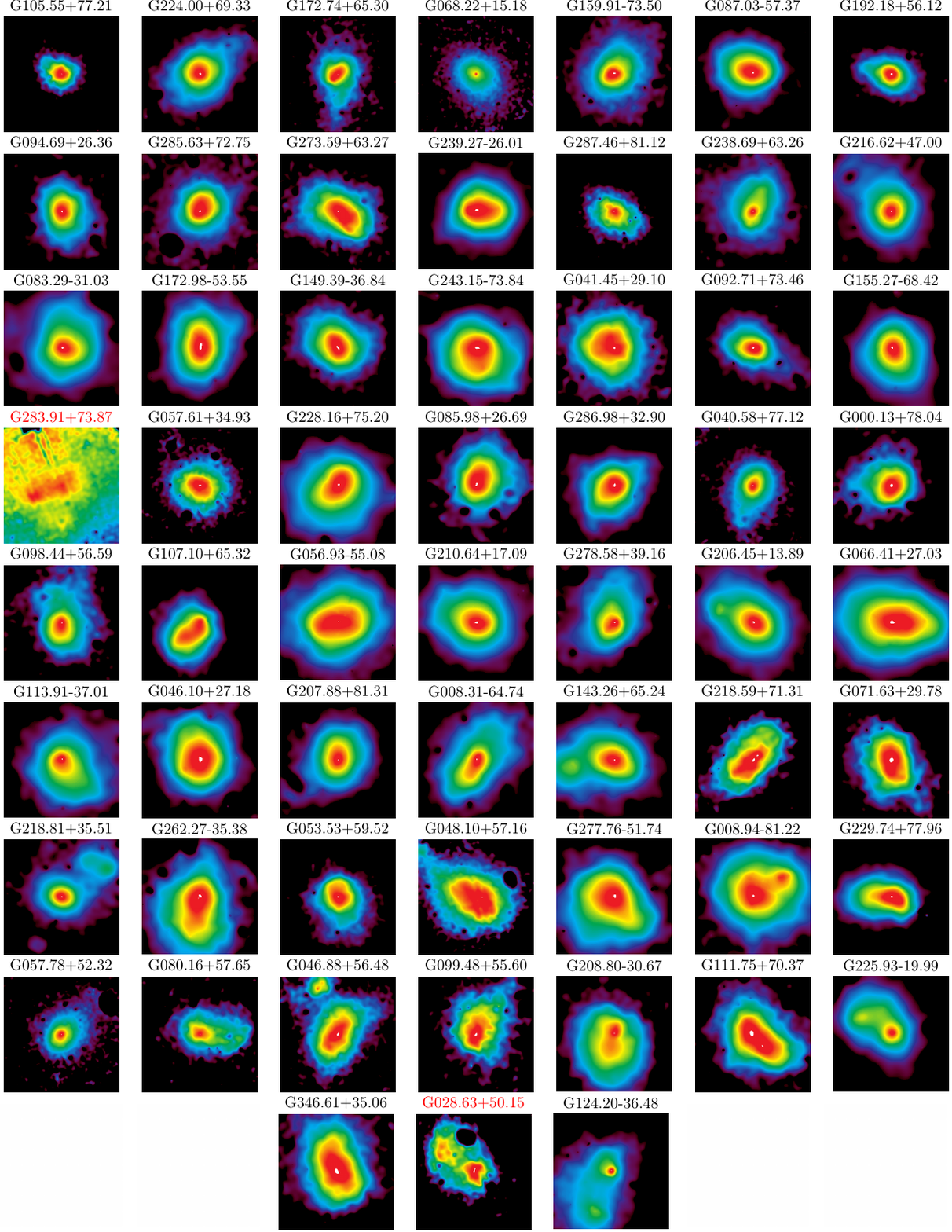}
        \caption{Continued.}
    \end{figure*}
    

\end{appendix}

\end{document}